\documentclass[journal,compsoc]{IEEEtran}
\usepackage{tikz}

\usepackage{xcolor}
\usepackage{wrapfig}
\usepackage{amsmath}
\usepackage{amsthm}
\usepackage{amssymb}
\usepackage{pifont}
\usepackage{multirow}
\usepackage{graphicx} 
\graphicspath{{./figures/}}
\usepackage{placeins}
\ifCLASSOPTIONcompsoc
\usepackage[caption=false,font=normalsize,labelfon
t=sf,textfont=sf]{subfig}
\else
\usepackage[caption=false,font=footnotesize]{subfi
g}
\fi

\usepackage{url}
\usepackage{booktabs} 
\usepackage{array} 
\usepackage{verbatim} 
\usepackage{enumitem}
\usepackage{cite}
\usepackage{bbm}

\usepackage{mathrsfs}

\usepackage[ruled,linesnumbered]{algorithm2e}
\usepackage{algorithmicx}
\usepackage{algpseudocode}

\usepackage[export]{adjustbox}
\usepackage[para]{threeparttable}
\usepackage{stfloats}

\DeclareMathOperator*{\argmin}{argmin} 
 
\usepackage{hyperref}

\begin{document}

\title{Quid pro Quo in Streaming Services:\\ Algorithms for Cooperative Recommendations}

\author{Dimitra Tsigkari, George Iosifidis, and Thrasyvoulos Spyropoulos
\thanks{*This manuscript has been accepted for publication in {\sc IEEE Transactions on Mobile Computing}, \href{https://ieeexplore.ieee.org/document/10026335}{DOI: 10.1109/TMC.2023.3240006}, date of acceptance: January 23, 2023.}
 \thanks{Dimitra Tsigkari and  George Iosifidis are with Delft University of Technology, Delft, Netherlands, email: d.tsigkari@tudelft.nl, g.iosifidis@tudelft.nl.}
\thanks{Thrasyvoulos Spyropoulos is with Technical University of Crete,  Chania, Greece, e-mail:
spyropoulos@tuc.gr}
\thanks{This work was conducted while Dimitra Tsigkari and Thrasyvoulos Spyropoulos were with Eurecom, Biot, France and it was supported by the French National Research Agency under the ``5C-for-5G'' JCJC project with reference number ANR-17-CE25-0001 and by the H2020 MonB5G project (grant agreement number 871780). This publication also emanated from research conducted with the financial support of the European Commission Grant No. 101017109 (DAEMON).}
\thanks{Part of this work appeared in the proceedings of the IEEE Global Communications Conference~(GLOBECOM) 2021~\cite{tsigkari2021globecom}.}
}

\IEEEtitleabstractindextext{%
\begin{abstract}
Recommendations are employed by Content Providers (CPs) of streaming services in order to boost user engagement and their revenues.  Recent works suggest that nudging recommendations towards cached items can reduce operational costs in the caching networks, e.g., Content Delivery Networks (CDNs) or edge cache providers in future wireless networks. However, cache-friendly recommendations could deviate from users’ tastes, and potentially affect the CP’s revenues. Motivated by real-world business models, this work identifies the misalignment of the financial goals of the CP and the caching network provider, and presents a network-economic framework for recommendations. We propose a cooperation mechanism leveraging the Nash bargaining solution that allows the two entities to jointly design the recommendation policy. We consider different problem instances that vary on the extent these entities are willing to share their cost and revenue models, and propose two cooperative policies, CCR and DCR, that allow them to make decisions in a centralized or distributed way. In both cases, our solution guarantees reaching a fair and Pareto optimal allocation of the cooperation gains. Moreover, we discuss the extension of our framework towards caching decisions. A wealth of numerical experiments in realistic scenarios show the policies lead to significant gains for both entities.
\end{abstract}
}

\maketitle

\section{Introduction}

\subsection{Background and Motivation}

Recommender systems~(RSs) permeate today's on-demand streaming services such as Netflix, Disney+, etc.; and are affecting substantially the content requests issued by their subscribers. In Netflix, for example, it is estimated that $80\%$ of the requests stem from the recommendations that are offered to its users~\cite{gomez2016netflix}. Indeed, by proposing contents that are relevant to their users' interests, Content Providers~(CPs) can increase the viewing activity in their platforms, reduce the user churn, and eventually boost their revenues~\cite{gomez2016netflix}. Therefore, it is not surprising that CPs comprehend the business value of these systems and invest research and financial resources to improve their accuracy.

At the same time, recommendations can be leveraged by content caching networks to steer user requests towards nearby-cached contents. These caching networks are either today's traditional Content Delivery Networks~(CDNs) or edge cache providers in future wireless architectures~(we will use, hereafter, the term CDN  to imply any such caching network provider).
The recently-coined terms of cache/network-friendly recommendations capture exactly this idea: recommendations aiming to reduce the CDNs' routing expenses without deviating irreparably from the users' viewing preferences. This is a  promising area of research with recent works proposing cache-aware recommendation policies, \emph{e.g.}, \cite{cache-centric-video-recommendation, kastanakis2020network}, and the joint optimization of caching and recommendation decisions, \emph{e.g.}, \cite{chatzieleftheriou2019joint-journal, tsigkari2022approximation, qi2018optimizing}. This idea not only can reduce the operating and retrieval costs of CDNs
but also can improve the service quality for the users by achieving smaller viewing start-up delays and/or higher bitrates of the streamed content~\cite{tsigkari2022approximation}. 

Clearly, RSs have already become a powerful tool affecting all key stakeholders in the content distribution ecosystem. And, as their  influence increases further, it is imperative to ensure they will foster synergies instead of creating misaligned incentives. Specifically, a hitherto unexplored aspect in this context is the \emph{tension} between CPs and CDNs when it comes to recommendations: the cache-friendly recommendations of CDNs may deviate from the users' interests and thus affect negatively the CPs' revenues; while the CPs' recommendations might induce costly data transfers for the CDNs. This problem is more pronounced in the case where  Over-The-Top (OTT) CPs  lease CDN infrastructure to deliver their services, but appears also in content streaming platforms with self-owned caching infrastructure.

The goal of this work is to investigate this new problem by: 1)~understanding and modeling the root causes of the CP's and CDN's potential conflicts  when it comes to recommendations; 2)~proposing a cooperation framework to enable their agreement; and 3)~designing algorithms for realizing this coordination based on the information the two entities want to disclose. The core of our proposal is the following simple and practical idea:  the CDN  charges lower content delivery fees to the CP when the latter agrees to tune its recommendations towards cached contents. This discount will balance the CP's expected viewing gains with the CDN's induced savings on retrieval costs. Devising these \emph{cooperative recommendations} is a new and highly non-trivial problem whose nature and complexity cannot be properly handled by existing approaches for cache-friendly recommendations. 
Such incentive-compatible recommendation policies have the potential to revolutionize streaming platforms, in the same way that the collaboration of ISPs and CDNs changed the scenery of content distribution, see~\cite{smaragdakis-padis} and references cited therein.

\subsection{Methodology and Contributions} \label{subsection:approach}
\begin{figure}[t]
	\centering  
	\includegraphics[width=8.9cm, trim={1.0cm 1.2cm 1.59cm 0.86cm},clip]{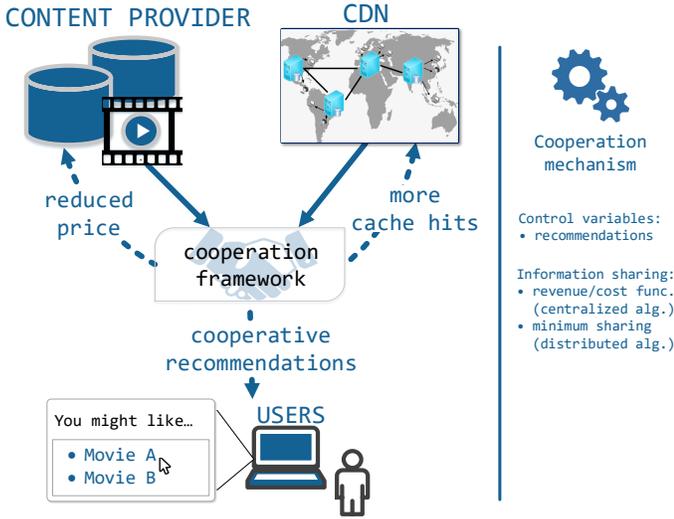} 
	\caption{The CP and CDN cooperate by  agreeing on the  recommendations the users receive. The incentives of the cooperation are provided by the reduced  price the CP is charged for the content delivery and the resulting increase in the number of cache hits that leads to lower retrieval costs for the CDN.}  
	\label{fig:collflow}
\end{figure}

Our proposal relies on a rigorous game-theoretic framework where we model the CP-CDN cooperation as a bargaining problem~\cite{nash1950bargaining}.  Our starting point and baseline will be a scenario where  the CP recommends contents based on its expected revenue~(and/or the users' interests) and the CDN makes caching decisions without any prior knowledge of the recommendations and how they shape content requests. 
On this basis, the CDN proposes to the CP a price discount  for delivering its contents, in exchange for tweaking the recommendations towards already-cached items~(see Fig.~\ref{fig:collflow}). In contrast to state-of-the-art cache-friendly recommendations or joint caching-recommendation schemes, this price discount provides a concrete incentive for the CP to adjust the recommendations the users receive.
This bargaining problem is formulated in a way that it 
 leads to a Pareto optimal and proportionally fair split of the cooperation gains, which is also incentive-compatible based on the Nash bargaining axioms.

To the best of our knowledge, this is the first work proposing the cooperation of the CP and CDN on the grounds of recommendations.  In summary, the contributions of this work are the following:
\begin{itemize}[leftmargin=*]
\item It identifies and models the new problem of misaligned incentives among the CP and CDN regarding the  recommendations offered to users. The employed system model is motivated by real-world business cases regarding the two entities' decision mechanisms and revenue models.

\item  It formulates a rigorous bargaining problem for addressing the trade-off between recommendation-induced revenues for the CP and retrieval costs for CDN in streaming services. The problem solution will allow them to devise the cooperative recommendations while splitting fairly the gains.
\item It proposes the Centralized Cooperative Recommendations~(CCR) algorithm for the scenario where the two entities share the necessary information regarding their  cost/revenue functions with a third party that solves the bargaining problem in a \emph{centralized} fashion.
\item It proposes the Distributed Cooperative Recommendations~(DCR) algorithm for the scenario where the CP and CDN have undisclosed private information. This leads to a \emph{distributed} bargaining solution where the CP and CDN solve their own problem instances while being oblivious to each other's private information. The two entities coordinate  through lightweight signaling that drives them eventually to the bargaining equilibrium.

\item It discusses how  the presented framework can be extended to cooperative caching policies and it analyses its difficulty. This problem of cooperative recommendations and caching turns out to be hard to solve but has the potential to further increase the cooperation gains.
\item Through a number of numerical evaluations using a real dataset and realistic system parameters, it verifies the efficacy and operation of the bargaining framework and explores the impact of key system parameters on the equilibrium properties. This provides rich insights on the potential economic benefits of our proposal and market design guidelines. 
\end{itemize}

\section{Problem Setup} \label{sec:setup}

\subsection{Recommendations, Content Requests and Caching}

In this work, we present a  cooperation scheme between the CP and CDN on the basis of the recommendations the former offers to its users. 
Following the current business models for the two entities,  we model their utility functions that represent their profit from the OTT market.

\textbf{Content Recommendation Model:} The CP owns a content catalog  $\mathcal{K}$ that is accessible to a set $\mathcal{U}$ of users through the CP's OTT service. In this work, we focus on catalogs of contents that are static, and thus cacheable. We will use the terms OTT or streaming services interchangeably to describe on-demand streaming services.  A (personalized) list of $N_u$ items are recommended to each user $u\in \mathcal{U}$. The recommendations are based on the predicted relevance of each content to the user's tastes, viewing history, context, etc. These relevances~(sometimes also called ``scores'' or ``rankings'') are calculated by today's state-of-the-art RSs~(that are employed by the CP) using techniques such as  collaborative filtering, deep neural networks, reinforcement learning, etc.~\cite{gomez2016netflix},\cite{davidson2010youtube}. We denote by $r_{ui}\in [0,1]$ these relevances.  Typically, the CP would select the $N_u$ items with the highest $r_{ui}$ or the highest expected revenue to feature the recommendations list of user $u$~\cite{gomez2016netflix},~\cite{tufekci2018youtube}. In this work, the recommendation decisions~(\emph{i.e.,} deciding which contents will appear in the user's recommendations list) are made not only based on the utilities $r_{ui}$ but also on the cooperation terms.
Our problem considers two sets of recommendations: 
\begin{itemize}[leftmargin=*]
\item \emph{(Input) Baseline Recommendations:} $Y^{b}=(y_{ui}^{b}\in\{0,1\}, u\in \mathcal{U}, i\in \mathcal{K})$, where $y_{ui}^{b}=1$ if content $i$ is recommended to user $u$. These are decided by the CP before any cooperation and are \emph{input parameters} for our problem. For example, these could be the top $N_u$ most relevant contents~(to each user), as mentioned above.
\item \emph{Cooperative Recommendation Variables:} $Y=(y_{ui}\in[0,1], u\in \mathcal{U}, i\in \mathcal{K})$, which are the \emph{probabilistic} recommendation variables optimized jointly by the CP and CDN. These are the \emph{control variables} of our problem.
\end{itemize}

\noindent Using continuous variables for the cooperative recommendations allows the CP to provide
some variety to the recommendations it offers to the same user
from session to session.

\textbf{Content Request Model:} Each user $u$ makes content requests according to the following model~ \cite{chatzieleftheriou2019joint-journal}, \cite{tsigkari2022approximation}:

\begin{itemize}[leftmargin=*]
    \item follows the recommendations with probability $\alpha_u$; where, w.l.o.g. each of the $N_u$ items is considered equally likely to be requested\footnote{The quantity $\alpha_u$ captures the percentage of time the user $u$ tends to follow the recommendations and can be based on user's past behavior.}. Hence, each recommended content is requested by the user with probability $\alpha_u/N_u$.
       \item with probability $(1-\alpha_u)$, the user ignores the recommendations and requests a content $i\in \mathcal{K}$ of the catalog with probability\footnote{The value of $p_{i}$ captures the probability that any user would request the content $i$ outside of recommendations ({\it e.g.}, through the search bar), and could relate to the aggregate interest in this content by users.} $p_{i}$.
    \end{itemize}

\textbf{Content Caching Model:}  A CP subscribes to a CDN provider through a Service Level Agreement (SLA) 
for the delivery of the contents to the users. 
The CDN manages a set of $C$ caches with capacity $\mathcal{C}_j$, $j=1, \ldots, C$.  
 Moreover, there is  a root cache $C_{0}$ 
that stores all the contents. 
We denote by $\sigma_i$ the size~(in Gb) of the content $i$ and we assume that $  \mathcal{C}_j \ll \sum_{i\in \mathcal{K}} \sigma_i$, as is common in most caching setups, \emph{e.g.,}~\cite{femto_JOURNAL2013}. The CDN optimizes the caching decisions based on performance (e.g., latency, cache hits) and cost criteria (routing costs).
These decisions are described as follows:

\emph{(Input) Baseline caching:} $X^{b}=(x_i^{b} \in \{0,1\}, i\in \mathcal{K}, j=1,\ldots,C)$ where $x_{ij}^{b}=1$ if content $i$ is fully stored in cache $j$. These are determined by the CDN before any cooperation and are input parameters.

To better focus on the mechanics of the cooperation, we will develop our framework in the context of cache-friendly recommendations, \emph{i.e.,} assuming that the caching policy is decided at a different timescale than the recommendations and is fixed during the cooperation.
 We revisit caching variables, and how these could potentially also be designed jointly with recommendations  later, in Sec.~\ref{sec:extension}.

\begin{table}[t] 
\caption{Important notation}
\label{notation_summary}
\centering 

\begin{tabular}{|c|l|}

\multicolumn{2}{@{}l}{\textbf{Content Requests and Recommendations}}\\
\hline
$\mathcal{K}$ & catalog of contents\\
\hline
$\mathcal{U}$ & set of users in the network\\
\hline

$N_u$ & number of recommended contents for the user $u$\\
\hline

$\alpha_u$ & probability that user $u$ follows the recommendations \\
\hline
 \multirow{2}{*}{$p_i$} & probability that a user requests content $i$ while \emph{not}\\
& following the recommendations \\
\hline
\multicolumn{2}{@{}l}{\textbf{Revenues and Costs}}\\
\hline
$\lambda$ & price per Gb requested that the CP pays to the CDN\\
\hline
 $\sigma_i$ & size of content $i$~(in Gb)\\
\hline
$\rho$ & discount on the delivery price $\lambda$, $\rho\in (0,1)$\\
\hline

$R_{ui}$ & CP's revenue (expected) from user $u$ for content $i$\\
\hline
$r_{ui}$ & relevance (predicted) of content $i$ to user $u$ \\
\hline
$K_{ui}$ & CDN's cost of delivering content $i$ to user $u$\\
\hline
\multirow{2}{*}{$U, \widetilde{U}$} & CP's and CDN's utility~(profit) functions respectively,\\
& $U^b, \widetilde{U}^b$ for baseline scheme~(pre-cooperation)\\
\hline
\multicolumn{2}{@{}l}{\textbf{Input Parameters}}\\
\hline
$y_{ui}^{b}$ & (input) baseline recomm., before any cooperation\\
\hline
$x_{ij}^{b}$ & (input) caching allocation, before any cooperation\\
\hline
\multicolumn{2}{@{}l}{\textbf{Variables}}\\
\hline
\multirow{2}{*}{$y_{ui}$} & cooperative recomm. variable corresponding to  \\
& user $u$ and content $i$  for the centralized solution \\
\hline
\multirow{2}{*}{$\psi_{ui}, \widetilde{\psi}_{ui}$} & cooperative recomm. variables for the distributed\\ & algorithm,
as decided by the CP and the CDN resp.\\
\hline
\end{tabular}
\end{table}

\subsection{Revenue/Cost Model and Utility Functions} 
\label{subsec:revenues_costs}
We will now consider the various sources of revenues and costs for the CP and CDN in order to define their utility functions. While these sources can, of course, be highly nuanced from scenario to scenario, we propose a model that tries to capture key elements while staying tractable.

\textbf{CP revenues:} When a user $u$ requests a content $i$, this content is associated with an expected revenue $R_{ui}$
that depends on the CP's revenue model~(ad-based, subscription-based, transaction-based, etc.~\cite{maz_revenue_models}) and  the associated costs related to the purchase of contents~(through licensing or production).
This information is estimated  by the CP in order to decide its pricing  strategy and is used as input for our model . For example, in the case of an ad-based revenue model, $R_{ui}$ can be estimated as a result of ad impressions that appear during content $i$.
Furthermore, this expected revenue depends on the content relevances $r_{ui}$ in a non-trivial way.  For this reason, we capture this relation by a fairly generic model:
\begin{equation}\label{eq:Rui}
R_{ui} = \phi_{ui}(r_{ui}),
\end{equation}
\noindent where $\phi_{ui}$ can be any nondecreasing function of $r_{ui}$ that describes the impact of user's (predicted) interest in a content on the CP's revenues\footnote{These functions are built by the CPs using historical data; and are typically concave capturing diminishing returns on the relevances $r_{ui}$.\label{foot:fui}}.
 For example, $\phi_{ui}$ could be related to the probability of a user abandoning the viewing session as a function of $r_{ui}$. 
 
\textbf{CP costs/CDN revenues:} The delivery of a requested content is made by the  CDN that charges the CP on a basis of the amount of transferred data~(as is the case in today's CDNs~\cite{cloudfront_pricing}). We remind the reader that these charges apply to CPs without an in-house CDN, which is still the case for a large number of CPs, \emph{e.g.,} Disney+, Hulu. 
 We assume that the CP has to pay $\lambda$ currency units per Gb requested\footnote{In order to capture different pricing schemes where the CDN charges the CP per Gb \emph{delivered} (and not only requested), the price $\lambda$  could be multiplied by  the probability of abandonment by the user.}. 

\textbf{CDN costs:}  The main source of expenditures for the CDN is the cost related to  the delivery of a requested content to the user. We let $\mathcal{C}(u)$  be the  subset of caches that  a user $u$ has access to including the root cache~(which is accessible by every user).  A request for content $i$ by user $u$ may be served by at least one of the small caches in $\mathcal{C}(u)$ where $i$ is stored.  If the content is not cached, it will be served by the root cache $C_0$.

We assume that every link between user $u$ and the caches in the  set $\mathcal{C}(u)$ is characterized by a delivery~(retrieval) cost~(for the CDN). We let $k_{uj}$ denote this cost per Gb for user $u$ by the cache $j$. 
The value of  $k_{uj}$ can be estimated as a result of transit fees the CDN pays to transit networks or Internet Service Providers~(ISPs)  to retrieve the content from the origin
servers of the CPs and make it available to the users.  Moreover, they can include maintenance-related costs, \emph{e.g.,} related to  storage capacity, hardware, estate, energy, etc~\cite{gourdin2017economics}. The delivery cost from the root cache $C_{0}$ to user $u$ is $k_{u0}$~(per Gb), where $k_{u0}> k_{uj}$ for all $j=1,\ldots,C$. 
The CDN serves each request through the lowest-cost cache that has  the requested item, as is common in most caching setups~\cite{femto_JOURNAL2013, borst2010}. Adopting the notation in~\cite{femto_JOURNAL2013}, we denote the sequence of increasing user-cache costs by $k_{u(1)},k_{u(2)},\ldots,k_{|\mathcal{C}(u)|}$. Then, based on the caching decisions $X^{b}$, the delivery cost for content $i$ by user $u$ is:
\begin{equation}\label{Kui_generic}
  K_{ui}(X^{b})=  \sum_{j=1}^{|\mathcal{C}(u)|} \left[  \sigma_i k_{u(j)} x_{i(j)}^{b}  \prod_{l=1}^{j-1} \left(1-x_{i(l)}^{b}\right) \right].
\end{equation}
According to the formula above, $x_{i(j)}^{b}  \prod_{l=1}^{j-1} \left(1-x_{i(l)}^{b}\right)$ will be equal to $1$ when the requested content $i$ is retrieved by the cache $(j)$, \emph{i.e.,} the  cache with the $j$-th  lowest user-cache cost, at cost $k_{u(j)}$,  for lack of any other cache with lower cost~($x_{i(l)}^b=0, l<j$). If $i$ is not cached
in any cache, it will be retrieved from the root cache $C_0$, which is ranked
last, resulting in high cost.

\textbf{CP's and CDN's utilities before cooperation:} Based on the above problem setup and revenue models, we can now derive the total \textit{utility} (revenues minus costs) each of the two parties enjoys before cooperating. We define the baseline~(initial)  utility of the CP before any cooperation  as the expected revenue minus the expected price  it has to pay to the CDN:
\begin{equation} \label{U_CP_initial}
U^{b}= \sum_{u\in \mathcal{U}} \sum_{i\in \mathcal{K}} \frac{\alpha_u}{N_u}  y^{b}_{ui} (R_{ui} -  \lambda \sigma_i ).
\end{equation}
We do not account for the revenue that
comes from the content requests that are not a result of
recommendations, i.e., when, with probability $1-\alpha_u$, the
user does not follow any of the recommendations. It is easy to see that these requests do not affect the cooperation~(whose control variables are the recommendations).
Moreover, note that the definition of  $U^{b}$ is generic and does not depend on how the CP devises the standard  recommendations~(\emph{i.e.,} the values $y_{ui}^{b}$).

Given the caching and recommendation decisions  before the CP-CDN cooperation, the baseline  utility of the CDN  is expressed as the expected revenue (from the delivery contract) minus the expected  delivery~(or retrieval) costs:

\begin{IEEEeqnarray}{lcl} \label{U_CDN_initial}
\widetilde{U}^{b}= &&  \sum_{u\in \mathcal{U}} \sum_{i\in \mathcal{K}}  \frac{\alpha_u}{N_u} y^{b}_{ui}  \big( \lambda   \sigma_i - K_{ui}\big). \IEEEeqnarraynumspace
\end{IEEEeqnarray}

\normalsize

\subsection{Towards Cooperative Decisions}
The goal of our cooperative framework is to improve the aforementioned utilities of \emph{both} parties. We are specifically interested to maximize the  gains and ensure they are ``fairly'' shared\footnote{For now, we use the words ``fair'' and ``unfair'' in an intuitive way: an entity that considers a solution~(\emph{i.e.,} the allocation of the gains)  unfair believes that this solution was achieved at the cost of this entity's own benefit. We will formally define and  elaborate on the fairness framework in Sec.~\ref{subs:NBS}.}. As explained earlier, such gains can result by motivating the CP to modify some of its original recommendations towards lower cost items (\emph{e.g.,} cached ones). To ensure that the CP will not lose revenue from these modifications (we remind the reader that this revenue relates to how related the recommended contents are for users, see \eqref{eq:Rui}) we assume the CDN offers a discount on the content delivery fees of such ``lower cost'' content. In particular, we let $\rho$ denote this normalized discount factor\footnote{This $\rho$ can be alternatively seen as a percentage discount on the price $\lambda$. So, for example, when $\rho=0.5$ or $50\%$, the CP would pay half the delivery price to the CDN for any modified recommendation.} on the  price  $\lambda$, where $0< \rho <1$. The value of $\rho$ is either set by the CDN or by a regulatory authority~(who acts as a mediator for their cooperation). We will discuss in Sec.~\ref{sec:perf} how the value of $\rho$ could be chosen in practice.
Then, the new price the CP would have to pay to the CDN is
\begin{eqnarray} \label{eq:lambda}
\Lambda_{ui} = \lambda    [ 1+ \rho (y_{ui}^{b} -1)].
\end{eqnarray}

\noindent Specifically, if a content $i$ is recommended now but it was not before the cooperation (\emph{i.e.}, $y_{ui}>0$ and $y_{ui}^{b}=0$), then the discount $\rho$ applies. If, on the contrary, the content continues to be recommended~(even partially) as before (\emph{i.e.}, $y_{ui}>0$ and $y^{b}_{ui}=1$), no discount applies.  
Our problem formulation, to follow, is applicable to either scenario, so w.l.o.g. we will focus on the former.
 We note that the requests that do not come through recommendations  are not subject to any discount.
Then, the new  utility functions for the CP and CDN are:

\begin{IEEEeqnarray}{rcl}
U&=&  \sum_{u\in \mathcal{U}} \sum_{i\in \mathcal{K}} \frac{\alpha_u}{N_u}  y_{ui} (R_{ui} -  \Lambda_{ui} \sigma_i),\IEEEeqnarraynumspace \label{U_CP} \\
\widetilde{U}&=&\sum_{u\in \mathcal{U}}  \sum_{i\in \mathcal{K}} \frac{\alpha_u}{N_u} y_{ui}\big( \Lambda_{ui}  \sigma_i - K_{ui}\big).\IEEEeqnarraynumspace \label{U_CDN}
\end{IEEEeqnarray}
\normalsize

\theoremstyle{remark} \newtheorem{remark:streaming}{Remark}
\begin{remark:streaming} \label{remark:streaming}
In line with related work on caching and recommendations policies, the proposed cooperation framework makes \emph{proactive} decisions~(on the recommendation variables). Although the presented framework deals with contents and not chunks of various qualities/bitrates, under small modifications, it can also treat files that correspond to pairs (content chunk, quality). Since online Adaptive Bitrate~(ABR) policies operate at a different timescale, \emph{i.e.,} during playback, we assume that they act in a complementary way on top of the proactive cooperation decisions. 
\end{remark:streaming}

\subsection{Toy Example} \label{subsec:toy}

\begin{figure}[bht]
	\centering  
	\includegraphics[width=9cm, trim={3.93cm 2.12cm 2.29cm 6.65cm},clip]{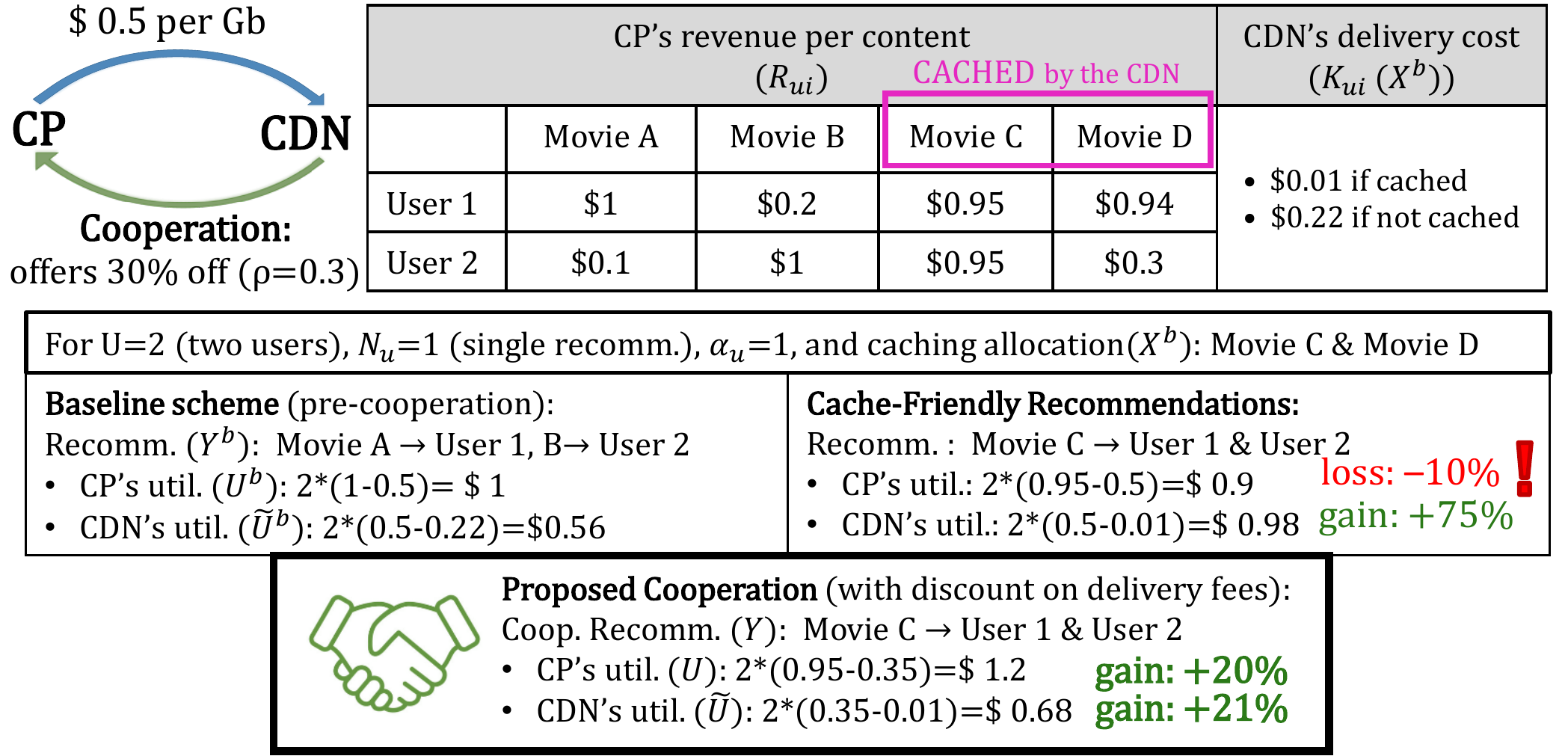} 
	\caption{Toy example presented in Sec.~\ref{subsec:toy}. In this example, the CP-CDN cooperation leads  to financial gains of $20\%$ and $21\%$ respectively. These gains derive from the discount on the delivery fees~(for the CP) and the fetching/retrieval savings~(for the CDN). In contrast, a typical cache-friendly recommendations approach~(without any discount on the delivery fees) would lead to a loss in profit for the CP, and a  gain in profit for the CDN~(that might be perceived as ``unfair'' by the CP).}
	\label{fig:toy}
\end{figure}

To better understand the cooperation model and the tradeoffs involved, we present a
toy example depicted in Fig.~\ref{fig:toy}. We consider a scenario with two users, a catalog of four equal-sized contents~(of $1$Gb size) and a single cache with capacity $2$Gb. Upon request, a content is served by the cache, if it is cached there. Otherwise, it will be served by the root cache.  For simplicity, we assume that the users will receive a single recommendation that will follow with probability $1$. 
The CP pays to the CDN $\$ 0.5$ per Gb~(outside of any cooperation) while the CDN offers to the CP a discount of $30\%$ on the delivery fees if they cooperate and the CP modifies its recommendations.  The CP's revenues per requested content and the CDN's costs related to the delivery of the contents are depicted in the table on the top right of the figure. The revenues $R_{ui}$ could be calculated, for example, as a result of ad impressions appearing during playback. We assume here that these revenues reflect  how relevant a content is to a user~(accounting for predicted abandonment rates), \emph{e.g.,} Movie A is the most relevant content to User 1, while Movies C and D are a bit less relevant for this user.

On the bottom of Fig.~\ref{fig:toy}, we see the recommendation decisions made in different scenarios, as well as the resulting utilities~(profits) of the two entities. In particular, outside of any cooperation~(baseline scheme), the CP would recommend the contents that will bring the highest revenue, \emph{i.e.,} Movie A to User 1 and Movie B to User 2, while  the CDN would cache some contents without knowledge of the recommendations and how they shape the requests. We assume that Movies C and D are cached based on the aggregate popularity observed in a period of time prior to the cooperation.  Therefore, the requests for the recommended contents will lead to cache misses and extra retrieval costs~(for the CDN).

On one hand, a typical cache-friendly or cache-aware recommendations policy, such as the ones in~\cite{cache-centric-video-recommendation, kastanakis2020network}, would recommend cached items that are still relevant to the users' tastes aiming for more cache hits. In our example, that would be recommending Movie C to both users. However, this would lead to a loss in profit for the CP~($-10\%$ when compared to the baseline scheme) and a large gain for the CDN~($+75\%$). Hence, the CP does not have concrete incentives in adjusting the recommendation towards cached items. Moreover, issues of trust, privacy, and coordination between the two entities could arise. We remind the reader that related works on the co-design of caching and recommendations~\cite{cache-centric-video-recommendation, kastanakis2020network, chatzieleftheriou2019joint-journal, tsigkari2022approximation, qi2018optimizing} trivially assume that both decisions are made by the same entity~(as is the case for only  a small number of OTT services, such as Netflix) and they do not explore the financial aspects of the recommendations.

On the other hand, if the two entities cooperate, incentives are provided to the CP (under the form of a discount on the delivery fees) and the resulting gains are split in a way that is not perceived as ``unfair'' by any party. The cooperative recommendations would suggest Movie C to both users, as it was the case with the cache-friendly recommendations. However, here,  the CP will pay reduced delivery fees that will compensate for the loss in its profit. At the same time, this cooperation is still profitable for the CDN who will avoid the extra delivery costs when compared to the baseline scheme. We see that, already in this toy example,  the cooperation leads to gains of at least $20\%$ for each entity. Note that any other solution, \emph{e.g.,} recommending Movie D to both users,  would result in worse gains for at least one entity, and thus in an ``unfair'' allocation of the cooperation gains.

In this example, it is easy to guess how to find the  cooperative recommendation policy that boosts both entities' profits. However, this task becomes significantly harder for bigger scenarios~(large content catalogs,
multiple recommendations per user, etc.). Moreover, one might wonder:  would the CP and CDN be willing to exchange information on their utility functions in order to find the solution~(since these functions constitute sensitive business information)? And how the cooperative recommendations can impact the users?
To this end,
in the next section, we formulate a cooperation mechanism while addressing these concerns.

\section{Problem Formulation and Algorithms} \label{sec:problem}

The toy example above illustrated that  the CP-CDN cooperation should provide incentives for both entities.  This means that the cooperative recommendation policy should satisfy: $U\geq U^{b}$ and $\widetilde{U}\geq \widetilde{U}^{b}$. 
As explained earlier, the CDN will propose a discount on the delivery fees in order to incentivize the CP to tune its recommendations towards cached contents.
Given this discount, the two parties~(or players, in game theory parlance) will try to benefit as follows:
\begin{itemize}
\item \emph{CDN:} it increases cache hits~(through cache-friendly recommendations) and thus it reduces the delivery costs~(term $K_{ui}$ in~\eqref{U_CDN}). These cost savings will compensate the lower delivery fees (term $\Lambda_{ui}$ in~\eqref{U_CDN}). 
\item \emph{CP:} it modifies the recommendations only if the cooperative ones lead to minor loses in expected revenue $R_{ui}$, that can be amortized by the applied fee reduction. 
\end{itemize} 
Moreover, both parties have the following concrete goals:  1)~benefit as much as possible~(hence the need for an \textit{optimization framework}), and 2)~reach an agreement that is perceived as \emph{fair} by both parties.

\subsection{Modeling the CP-CDN cooperation as a Nash Bargaining Solution}
\label{subs:NBS}

Having these desired properties as guideline, we model  the cooperative recommendations problem as a Nash Bargaining Solution~(NBS)~\cite{nash1950bargaining, myerson2013game}~from the cooperative game theory. 
The NBS is defined as the maximization of the product of payoffs~(\emph{i.e.,} the utility gains) of the two entities subject to individual rationality constraints, or equivalently the maximization of the logarithm of this product where the constraints are implicit in the domain of the logarithms~\cite{myerson2013game}. Therefore, the NBS would be
\begin{equation} \label{eq:NBS_general}
\max_{Y} \left[\log(U(Y) - U^{b}) + \log(\widetilde{U}(Y)- \widetilde{U}^{b})\right],
\end{equation}
where $U(Y) - U^{b}$ and  $\widetilde{U}(Y)- \widetilde{U}^{b}$ represent the gains in utility of the CP and the CDN from a potential cooperation.

By formulating our problem in this way, the solution uniquely satisfies the Nash's axioms~\cite{nash1950bargaining, myerson2013game}. First, the solution is Pareto optimal, that is, there is no other solution that would benefit one party more without deteriorating the
other party's gains.  We also provide  the formal definition of Pareto optimality below. For this,  we use the following notation: for two vectors $(a_1,a_2)\in \mathbb{R}^2$ and $(b_1,b_2)\in \mathbb{R}^2$, the notation $(a_1, a_2) \geq (b_1, b_2)$ implies that $a_i \geq b_i$ for $i=1,2$. 
\theoremstyle{definition} \newtheorem{def:pareto}{Definition}
\begin{def:pareto}[Pareto optimality, from~\cite{bertsimas2011price}]
A point $A$ in the feasible set $\mathcal{F}$~(of recommendation policies) is Pareto optimal~(or strongly Pareto efficient) if there is no other point $B$ in $\mathcal{F}$ such that $(U(B), \widetilde{U}(B)) \geq (U(A), \widetilde{U}(A))$ and $(U(B), \widetilde{U}(B)) \neq (U(A), \widetilde{U}(A))$.
\end{def:pareto}

Imagine, for example, that  there are only two feasible recommendation policies $A$ and $B$ where $A$ leads to gains of  $25\%$ for the CP  and $30\%$ for the CDN, while $B$ leads to gains of  of $10\%$ and  $30\%$ respectively. Then, for this problem instance, the NBS would yield as solution the policy $A$, which is Pareto optimal.

Another property that is guaranteed by the NBS is proportional fairness. In other words, the solution of the NBS is such that, when compared to any  other feasible allocation of gains, the aggregate proportional change in utilities is less than or equal to zero. We provide below the formal definition:
\theoremstyle{definition} \newtheorem{def:PF}[def:pareto]{Definition}
\begin{def:PF}[Proportional Fairness, from \cite{bertsimas2011price}]
A point $A$ in the feasible set $\mathcal{F}$~(of recommendation policies)   is proportional fair  if for any  other point $B$ in $\mathcal{F}$ the following is true:
\begin{equation*}
\frac{U(B)- U(A)}{U(A)}+ \frac{\widetilde{U}(B)- \widetilde{U}(A)}{\widetilde{U}(A)} \leq 0.
\end{equation*}
\end{def:PF}

\noindent In our setting, a (cooperative) recommendation policy is considered proportional fair if any other policy would lead to a percentage decrease in utility of one entity that is larger than the percentage increase of the other entity. Imagine, for example, that  the CP-CDN cooperation would yield  only two possible recommendation policies $A$ and $B$ where $A$ leads to gains of  $25\%$ for the CP  and $30\%$ for the CDN, while $B$ leads to gains of  of $50\%$ and  $25\%$ respectively. Then, for this problem instance,  policy $B$ is the proportional fair solution.  We refer the reader to~\cite{bertsimas2011price} for a broad discussion on fairness.

Furthermore, due to the implicit domain constraints deriving from~\eqref{eq:NBS_general}, \emph{i.e.,} $U-U^{b} \geq0$ and $\widetilde{U}- \widetilde{U}^{b} \geq0 $, the payoff of every entity is no worse than the payoff it would get outside of any cooperation, \emph{i.e.}, $(U^{b}, \widetilde{U}^{b})$.  In fact,  ($U^{b}, \widetilde{U}^{b})$ is the ``disagreement point'' of the cooperation~\cite{nash1950bargaining}: if  $U < U^{b}$ or $\widetilde{U} < \widetilde{U}^{b}$, there will be no feasible solution and, thus, no agreement on cooperation.
Therefore, both parties have an incentive to cooperate. 
Moreover, if the positions of the two entities~(in terms of utility functions and the disagreement point) are  symmetric, then the solution treats them symmetrically.

In Sec.~\ref{subs:centralized}, we formulate  in detail the problem that would allow the CP and the CDN to devise the cooperative recommendations policy in a centralized way, where the two entities share all the necessary information for its solution. In Sec.~\ref{subsec:distributed},  we formulate the problem where the two entities exchange minimal information and we propose a decentralized algorithm that allows them to decide on the cooperative recommendations.

\subsection{Centralized Cooperative Recommendations}
\label{subs:centralized}

We will first formulate and study the centralized problem where the two entities share their cost/revenue functions.

\theoremstyle{definition} \newtheorem*{CEN-cache-single}{CCR: Centralized Cooperative Recommendations}
\begin{CEN-cache-single} 
\begin{IEEEeqnarray}{rclll} 
&\underset{Y}{\text{ min }}& \bigg[-\log\bigg( \sum_{u, i} \frac{\alpha_u}{N_u}  &y_{ui} (R_{ui} -  \Lambda_{ui} \sigma_i)- U^{b}\bigg)&\nonumber \IEEEeqnarraynumspace \\ 
&&\ \; - \log\bigg(\sum_{u, i} \frac{\alpha_u}{N_u} &y_{ui}( \Lambda_{ui}  \sigma_i - K_{ui}) - \widetilde{U}^{b}\bigg)&\bigg]\IEEEeqnarraynumspace \label{ccr:objective}\\
&\text{s.t.  }&  \sum_{i \in \mathcal{K}} y_{ui} = N_u, \; &\forall  u\in \mathcal{U},& \label{N_recomm}\\
&& y_{ui} \in [0,1], \; &\forall  u\in\mathcal{U}, i\in\mathcal{K},& \label{continuous_x_y}
\end{IEEEeqnarray}
\end{CEN-cache-single}
\noindent where the baseline utilities $U^{b}$ and $\widetilde{U}^{b}$ are defined in~\eqref{U_CP_initial} and~\eqref{U_CDN_initial}. The constraints in~\eqref{N_recomm} suggest that each user receives $N_u$ recommendations\footnote{If the solution $Y^*$ contains more than $N_u$ positive values~(due to the probabilistic model) we can easily select exactly $N_u$ following the technique in~\cite{blaszczyszyn2015optimal} and being compatible with $Y^*$ on expectation.}.

Moreover, we note that the inequalities $U-U^{b} =\sum_{u, i} \frac{\alpha_u}{N_u}  y_{ui} (R_{ui} -  \Lambda_{ui})-U^{b} \geq0$ and $\widetilde{U}- \widetilde{U}^{b} = \sum_{u, i} \frac{\alpha_u}{N_u} y_{ui}( \Lambda_{ui}   - K_{ui})-\widetilde{U}^{b} \geq0 $
are implicit constraints as the domain of the logarithms must be non-negative. 
Since   ($U^{b}, \widetilde{U}^{b})$ is the disagreement point,   if  $U < U^{b}$ or $\widetilde{U} < \widetilde{U}^{b}$, there will be   no agreement on cooperation, by definition.  Then the CP will keep its baseline recommendations $Y^{b}$ while the CDN will not provide a price discount. 
The next lemma shows that the CCR problem is tractable.
\theoremstyle{plain} \newtheorem{lemma:biconvex}{Lemma} 
\begin{lemma:biconvex} \label{lemma:biconvex}
The CCR Problem is (strictly) convex.
\end{lemma:biconvex}
\begin{proof}
The objective function is (strictly) convex since the logarithm is a concave function and the arguments of the logarithms are linear functions of $Y$. Moreover, the problem's constraints are linear.
\end{proof}

As a result of Lemma~\ref{lemma:biconvex}, standard
interior-point or dual methods would efficiently give the unique optimal solution. We summarize below the algorithm to devise the cooperative recommendations in a centralized manner. 

\theoremstyle{definition} \newtheorem*{CCR_alg_def}{The CCR Algorithm}
\begin{CCR_alg_def}
The CP communicates the values $\alpha_u$, $N_u$, $Y^{b}$,  $U^{b}$ and its utility function $U$. 
The CDN communicates the discount $\rho$, the value of $\widetilde{U}^{b}$, and its utility function $\widetilde{U}$. 
Then,  the CCR Problem is solved through standard
 dual or interior-point methods. It returns $Y^*$, the optimal cooperative recommendation policy.
\end{CCR_alg_def}

The process described above could be managed either by a trusted third party~(cooperation mediator) that collects the necessary information or by the two entities together. Given that today's major  CDNs update/fill their caches during off-peak hours, as is the case with Netflix's CDN~\cite{netflix_and_fill}, the algorithm could run at any time after the fill window, and it could concern the expected requests in the period of time until the next cache update or in a period of a few hours.

\theoremstyle{remark} \newtheorem{remark:https}[remark:streaming]{Remark}
\begin{remark:https} \label{remark:https}
The proliferation of encrypted user-cache communication through HTTPS/TLS requests is considered an obstacle for efficient content caching~(and, thus, cache-friendly recommendations) within the OTT services. However,  there are protocols proposed in literature that can ensure that the CDN’s caches are blind to the cached contents~(\emph{e.g.,}~\cite{eriksson2017blind}). Similar protocols could be embedded in our
framework since the CDN needs only to estimate the retrieval cost of the cached items~(that could be encrypted). Designing such a protocol is an interesting direction for future work but it goes beyond the scope of this manuscript.
\end{remark:https}

\subsection{Distributed Cooperative Recommendations with Minimal Information Sharing}\label{subsec:distributed}

As explained earlier, in order to solve the CCR Problem the two entities need to share their utility functions. However, in the highly competitive ecosystem of streaming services and content distribution, these functions constitute sensitive information.  Withholding such information might prevent the two parties from cooperating. Therefore, there is need for a cooperation mechanism that can assure privacy. Establishing such a mechanism  is not trivial since fairness~(along with the other properties of NBS) needs to be guaranteed, as in the centralized solution. We remind the reader that the NBS framework requires that both utility functions are taken into account in the same objective~(see~\eqref{eq:NBS_general}).

We overcome this challenge by applying the \emph{Alternating Direction Method of Multipliers}~(ADMM)~\cite{boyd2011distributed} to solve the problem in a distributed way. The idea behind ADMM is to \emph{split} the problem into two subproblems, where each subproblem contains only one entity's utility function. Then, the cooperative recommendation problem  is solved iteratively: each entity solves the subproblem that contains only its utility function and finds its local solution. Through coordination and after a sufficient number of iterations,  the subproblems' solutions coincide. The coordination preserves the entities' private information and is carried out by  a cooperation mediator, which is  either a trusted third party or the two entities together.
In order to define this distributed algorithm, 
in what follows:  1)~we reformulate the CCR Problem into an equivalent problem~(DCR Problem) that can be split into two subproblems, 2)~based on the theory on ADMM, we propose the distributed DCR algorithm, and 3)~we prove that the resulting cooperation gains converge to the ones of the centralized problem.

 Instead of the recommendation variables $Y$, we introduce here the local recommendation variables $\Psi=(\psi_{ui}\in [0,1])$ and $\widetilde{\Psi}=(\widetilde{\psi}_{ui}\in [0,1])$ that are the variables in the CP's and CDN's subproblems respectively. We reformulate the CCR Problem into the following equivalent problem: 
 
\theoremstyle{definition} \newtheorem*{cache-single}{DCR: Distributed Cooperative Recommendations}
\begin{cache-single} 
\begin{IEEEeqnarray}{rclll} 
&\underset{\Psi, \widetilde{\Psi}}{\text{ min  }}& \bigg[-\log\bigg( \sum_{u, i} \frac{\alpha_u}{N_u}  &\psi_{ui} (R_{ui} -  \Lambda_{ui} \sigma_i)- U^{b}\bigg)&\nonumber \IEEEeqnarraynumspace \\ 
&&\ \; - \log\bigg(\sum_{u, i} \frac{\alpha_u}{N_u} &\widetilde{\psi}_{ui}( \Lambda_{ui} \sigma_i  - K_{ui}) - \widetilde{U}^{b}\bigg)&\bigg]\IEEEeqnarraynumspace \\ 
&\text{s.t.  }&  \psi_{ui} = \widetilde{\psi}_{ui}, &\forall u\in\mathcal{U}, i\in\mathcal{K},& \label{dcr:consitency_con}\\
&& \sum_{i \in \mathcal{K}} \psi_{ui} = N_u,  &\forall u\in \mathcal{U},& \label{dcr:number_rec_1}\\
&& \sum_{i \in \mathcal{K}} \widetilde{\psi}_{ui} = N_u, &\forall u\in \mathcal{U},& \label{dcr:number_rec_2}\\
&& \psi_{ui}, \widetilde{\psi}_{ui}\in [0,1], &\forall u\in\mathcal{U}, i\in\mathcal{K},& \label{dcr:domain_constr} 
\end{IEEEeqnarray}
\end{cache-single}
\noindent where $\psi_{ui}$ and $\widetilde{\psi}_{ui}$ are the local recommendation variables as decided by the CP and the CDN respectively.  The constraints in~\eqref{dcr:consitency_con} are the \emph{consistency constraints} that require all  local recommendation variables to agree.

The augmented Lagrangian for the DCR problem is:
\begin{IEEEeqnarray}{rcl}
L_q(\Psi, \widetilde{\Psi}, Z) &=& -\log\left( U(\Psi)- U^{b}\right) - \log\left(\widetilde{U} (\widetilde{\Psi})-\widetilde{U}\right) \nonumber\IEEEeqnarraynumspace\\
&+&\sum_{u,i} z_{ui} (\psi_{ui} - \widetilde{\psi}_{ui}) + \frac{q}{2} \big|\big|\Psi - \widetilde{\Psi}\big|\big|_{F}^2, \label{eq:dcr:lagra}
\end{IEEEeqnarray} 
\normalsize
\noindent where $Z=(z_{ui})$ are the dual variables, $q$ is the penalty parameter, and $||\cdot||_{F}$ the Frobenius norm.  We remind the reader that the Frobenious norm of a matrix is defined as the square root of the sum of the squares of the matrix's entries. Moreover, the dual function is
\begin{equation}
d(Z) = \inf_{\substack{(\Psi, \widetilde{\Psi})\\ \text{s.t.~\eqref{dcr:number_rec_1}-\eqref{dcr:domain_constr}}}} L_q(\Psi, \widetilde{\Psi}, Z)
\end{equation}
\normalsize

\begin{figure}[tb]
	\centering  
	\includegraphics[width=9.14cm, trim={4.87cm 6.26cm 4.04cm 4.93cm},clip]{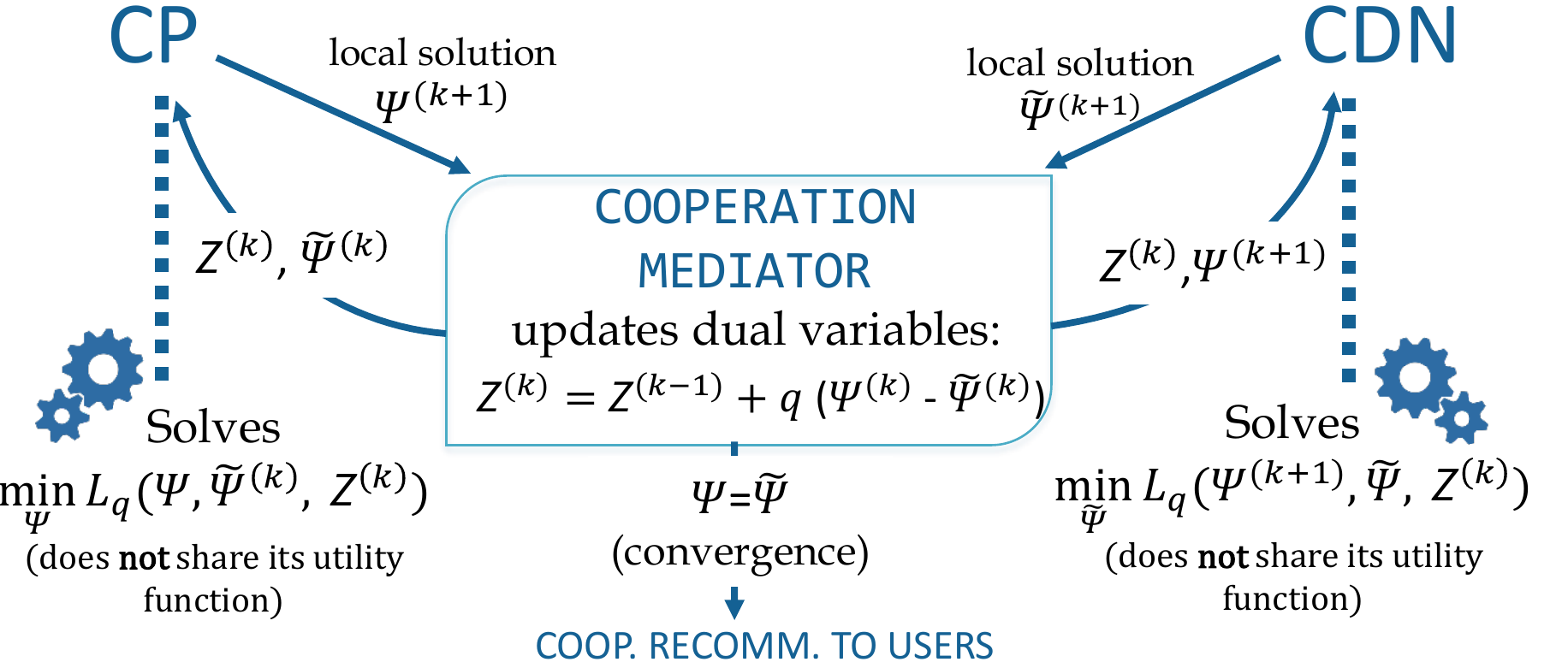} 
	\caption{Illustration of the DCR algorithm's steps. Each entity solves its subproblem~(without sharing its utility function) based on the other's local solution and the dual variables. They communicate  their local solutions to the cooperation mediator that updates and communicates the dual variables.}
	\label{fig:distributedalg}
\end{figure}
The ADMM for the DCR Problem is described below~(see also Fig.~\ref{fig:distributedalg}):

\theoremstyle{definition} \newtheorem*{DCR_alg_def}{The DCR algorithm}
\begin{DCR_alg_def}
The CP communicates $\alpha_u$, $N_u$ , $Y^{b}$ and the value of $U^{b}$. 
The CDN communicates $\rho$ and the value of $\widetilde{U}^{b}$.
Then, at every iteration $k+1$:
\begin{itemize}
\item The CP solves its subproblem and communicates its local solution:
\begin{equation}
\Psi^{(k +1)} := \argmin_{\substack{\Psi \\ \text{s.t.~\eqref{dcr:number_rec_1}, \eqref{dcr:domain_constr}}}} L_q \left(\Psi, \widetilde{\Psi}^{(k)}, Z^{(k)}\right). \label{eq:dcr:cp_sub}
\end{equation}
\item The CDN solves its subproblem and communicates its local solution:
\begin{equation}
\widetilde{\Psi}^{(k +1)} := \argmin_{\substack{\widetilde{\Psi}\\ \text{s.t.~\eqref{dcr:number_rec_2},\eqref{dcr:domain_constr}}}} L_q \left(\Psi^{(k+1)}, \widetilde{\Psi}, Z^{(k)}\right).\label{eq:dcr:cdn_sub}
\end{equation}
\item The cooperation mediator updates and communicates the dual variables:
\begin{equation}
Z^{(k+1)} := Z^{(k)} + q  \left(\Psi^{(k+1)} - \widetilde{\Psi}^{(k+1)}\right). 	 \label{eq:update_dual_variables}
\end{equation}
\end{itemize}
\end{DCR_alg_def}

We highlight here that each entity keeps private its utility function from the other entity and the mediator. 
The two entities reveal only their local solutions~($\Psi^{(k +1)}$ and $\widetilde{\Psi}^{(k +1)}$) at every iteration. These matrices are often sparse  leading to a low communication overhead at every iteration. This coordination until convergence~(that could involve only a few iterations) will occur at the beginning of the cooperation window.
Concerning the practicalities of the DCR algorithm, its iterations will terminate according to standard residual criteria~(see~\cite{boyd2011distributed}). We note that ADMM  tolerates  inexact minimization for the subproblems under the condition that the relative errors are summable~\cite{eckstein1992douglas}. Moreover, when the subproblems are solved in an iterative way, the warm-start technique can speed up the process. 
The following lemma  guarantees that the DCR algorithm converges~(after a sufficient number of iterations) to the centralized objective function value and solution. 

\theoremstyle{plain} \newtheorem{lemma:admm}[lemma:biconvex]{Lemma} 
\begin{lemma:admm} \label{lemma:admm}
If $p^*$ is the optimal value of the CCR Problem, and $DO^{(k)}$ is the DCR Problem's objective function value at iteration $k$, i.e., $DO^{(k)} = -\log\big(U(\Psi^{(k)})- U^{b}\big)-\log\big(\widetilde{U}(\widetilde{\Psi}^{(k)})-\widetilde{U}^{b}\big)$, then $ DO^{(k)} \rightarrow p^*$, as $k \rightarrow \infty$. Moreover, if $Y^*$ is the (unique) solution of the CCR Problem, then $\Psi^{(k)}, \widetilde{\Psi}^{(k)} \rightarrow Y^*$, as $k \rightarrow \infty$.
\end{lemma:admm}

\begin{proof}
According to the results in~\cite{boyd2011distributed}, we need to prove two conditions: 1) the extended-real-valued functions $-\log(U(\Psi)-U^{b})$ and $-\log(\widetilde{U}(\widetilde{\Psi}) - \widetilde{U}^{b})$  are closed, proper, and convex, and 2) the unaugmented Lagrangian $L_0$ has a saddle point. 
 The two functions are indeed convex~(in fact strictly convex) and closed. The corresponding extended-real-valued functions are proper since they are not identically equal to $+\infty$. 
We will now prove that  strong duality holds. When a feasible primal solution $\Psi^*=\widetilde{\Psi}^*$ exists such that $U(\Psi^*)-U^{b} >0 $ and $\widetilde{U}(\Psi^*)-U^{b} >0 $, then strong duality holds by Slater's condition~(which reduces to feasibility when the problem constraints are linear). 
Therefore, by feasibility and by strong duality, it follows that the unaugmented Lagrangian $L_0$ has a saddle point~\cite{boyd2004convex}.

Since we proved objective convergence, then the local solutions will converge to the unique centralized solution $Y^*$ since the DCR's objective function is strictly convex. This means that there is at most one global minimizer. 
\end{proof}

Essentially, Lemma~\ref{lemma:admm} implies that the properties of the centralized solution inherited by the NBS framework~(Nash axioms, see Sec.~\ref{subs:centralized}) hold also for the DCR's solution. This is important since it guarantees that the cooperation gains and the fair split of these gains will not be compromised when the two entities apply the DCR algorithm~(instead of the CCR). Finally, in Sec.~\ref{sec:perf}, we will see in practice how the convergence to the solution of the CCR problem is achieved as a function of the number of iterations, and how  the two entities' gains are  affected from iteration to iteration.

\subsection{Recommendations of high quality for the users}

One might argue that the cooperative recommendations could have potentially an impact on the users by degrading the recommendations they receive. Note that the user's interest in the content is one of the factors that determine the user's overall experience in OTT services, as shown in experiments~\cite{li2016impact}. However, the CP can limit a potential recommendation degradation by adding extra constraints in the problem. For example, adding in the CCR Problem the constraints 
\begin{equation} \label{eq:qor_constraint}
\sum_i \frac{y_{ui} r_{ui}}{N_u} \geq T_u, \text{ for every user } u,
\end{equation}
forces the average relevance of the cooperative recommendations to the user $u$ to be at least equal to a threshold $T_u\in(0,1]$. Adding these constraints does not have an impact on the problem analysis~(since they are linear with respect to the variables $Y$). In the (distributed) DCR Problem, the same constraints (with the local variables $\psi_{ui}$ instead of $y_{ui}$) can be applied when the CP solves its subproblem~(with no need of communicating these constraints to the CDN).

There is a common misconception that cache-friendly recommendations concern only a few~(very) popular contents. Although it has been shown that there exist a popularity bias in the core of RSs~(see~\cite{cremonesi2010performance, abdollahpouri2019unfairness}), related work on cache-friendly recommendations imposes  constraints similar to~\eqref{eq:qor_constraint} in order to better match the users' tastes beyond the universally popular contents. Going one step further, \emph{joint} caching and recommendation policies have been shown to outperform naive policies that would cache and recommend the most popular contents~(contents with highest aggregate popularity)~\cite{tsigkari2022approximation}. In fact, the joint approach yields a more efficient caching allocation and higher quality of recommendations since it makes decisions based on the diverse users' tastes and similarity between contents~(in terms of relevance to the users).
For this reason, in the next section, we extend the cooperation to the caching decisions.

\section{Extension to caching decisions}\label{sec:extension}

So far, we have focused our framework on scenarios where the recommendations are the only variables that can be re-designed by the CP and CDN, in the timescale of interest. A natural question that arises is whether also modifying the caching decisions in parallel, could yield even better profits: recommendations could concern contents that are cached in the cache that is closest to the user while they still bring high revenue to the CP.
 This is particularly useful in today's and future wireless architectures where caches are small while the CP's catalog is constantly growing. This is also in line with recent works proposing  the joint optimization of caching and recommendation decisions, \emph{e.g.}, \cite{chatzieleftheriou2019joint-journal}, \cite{tsigkari2022approximation}. Nevertheless, none of these works either explores the financial aspects of the caching-recommendation interplay, nor is it straightforward how to include these into our problem formulation and solution methodology.

A complete treatment of this topic goes  beyond the scope of this manuscript, due to the additional complexity it introduces in the solution methodology, and is subject to future work. Nevertheless, we will show here how to  include such variables into our model, and provide some preliminary analysis and a heuristic for this extended problem. We complement this analysis with related validation results in Sec.~\ref{sec:perf} that already show the proposed method can further increase the cooperation gains for both parties.

\textbf{Caching Setup and Variables.} In this section, we consider, for simplicity, a scenario where the CDN manages only one small cache whose capacity is $\mathcal{C}_1$.  Moreover, there is  a root cache $C_{0}$ 
that stores all the contents. We employ the prevalent continuous caching model that is valid either when coded caching is used~\cite{femto_JOURNAL2013} or when the files can be divided in equally-sized chunks and stored independently~\cite{blaszczyszyn2015optimal},~\cite{Paschos-infocom2016}. 
Therefore, for simplicity, in this section, we assume that the contents are equal-sized~(divided in chunks)\footnote{This assumption could be removed while our analysis and solution method could still be applied. However, in the case where contents are of heterogeneous sizes and  $x_i$ is interpreted as caching probability, the cache capacity constraint~(that we will formulate in what follows) will be satisfied in expectation.}. In addition to the variables and input values that were introduced in Sec.~\ref{sec:setup}, we define the cooperative caching variables:

\textit{Cooperative caching variables:} $X=(x_{i}\in [0,1], i\in \mathcal{K})$, where $x_{i}$ is the portion of content $i$ that is stored at the cache or the probability that the content $i$ is cached. These~(together with the recommendation variables $y_{ui}$) constitute control variables. 

We optimize proactive caching decisions, which constitute a key element of CDN's operations today, as explained before.  Therefore, as is common in related work~\cite{femto_JOURNAL2013}, we assume that the CDN proactively stores contents in its caches and this allocation stays fixed during the period between two updates/fills and between two CP-CDN cooperation instances.

Similar to~\eqref{Kui_generic}, the retrieval cost for the CDN  in a single cache~(and the root cache $C_0$) is:
\begin{equation}
K_{ui}(X) = k_{u0} + x_i(k_{u1}-k_{u0}),
\end{equation}
\emph{i.e.,} the cost will be $k_{u1}$ if the content $i$ is entirely cached in the (small) cache~($x_i=1$), $k_{u0}$ if it is not cached ~($x_i=0$), and a sum of the portions of these two costs otherwise.
In contrast to the definition of the utility functions in Sec.~\ref{sec:setup}, we  redefine here the CDN's utility function in order to include the profit that comes from requests out of recommendations~(note that now this term contains the control variables $x_i$). The CDN's baseline utility ~(before cooperation) and the utility for cooperation are defined as:
\begin{IEEEeqnarray}{rcl} 
V^{b}& = &\widetilde{U}^{b} + \sum_{u,i}(1-\alpha_u) p_i \left(\lambda - k_{u0} - x_i^{b}(k_{u1}-k_{u0})\right) \IEEEeqnarraynumspace \label{V_initial}\\
V &=&  \sum_{u,i}\Big[\frac{\alpha_u}{N_u} y_{ui}\left( \Lambda_{ui}   - k_{u0} - x_i(k_{u1}-k_{u0})\right)
\IEEEeqnarraynumspace \nonumber\\
 &&\quad \quad \; +(1-\alpha_u) p_i \left(\lambda - k_{u0} - x_i\left(k_{u1}-k_{u0}\right)\right)\Big]. \IEEEeqnarraynumspace \label{V_CDN}
\end{IEEEeqnarray}
In the second summand of $V$, the delivery fees are $\lambda$ since the discount does not apply to requests out of recommendations. We stress here that, for the CP, the corresponding term does not contain any of the control variables and it cancels out in the difference $U-U^{b}$. We can now formulate the optimization problem that can allow us to devise cooperative recommendation and caching policies in a centralized\footnote{Due to space constraints, we only present the centralized problem here. In fact, the presented framework could be also implemented in a distributed way.} manner~(with information sharing between the two entities).

\theoremstyle{definition} \newtheorem*{CCRcache}{CCRCache: Centralized Cooperative Recommendations \& Caching}
\begin{CCRcache} 
\begin{IEEEeqnarray}{rclll} 
&\underset{X,Y}{\text{ min }}&  \left[-\log \left(U(Y) - U^{b}\right)- \log\left( V(X, Y) - V^{b} \right)\right] \label{ccrcache:obj} \IEEEeqnarraynumspace \\
&\text{s.t.}&  \sum_{i \in \mathcal{K}} y_{ui} = N_u,\;\forall u\in \mathcal{U}, \label{ccrcache:N_recomm}\\
&& \sum_{i \in \mathcal{K}} x_i \leq C,  \label{ccrcache:capacity_con}\\
&& x_i, y_{ui} \in [0,1], \;\forall u\in\mathcal{U}, i\in\mathcal{K}, \label{ccrcache:continuous_x_y}
\end{IEEEeqnarray}

\end{CCRcache}
\normalsize
\noindent where $U^{b}$ and $V^{b}$ are in defined in~\eqref{U_CP_initial} and \eqref{V_initial}. According to~\eqref{V_initial} and~\eqref{V_CDN}, the CDN's gain in utility is:
\small
\begin{IEEEeqnarray}{rcl} 
V(X, Y)- V^{b} =&& \sum_{u, i} \big[\frac{\alpha_u}{N_u} y_{ui}\big( \Lambda_{ui}   - k_{u0} - x_i(k_{u1}-k_{u0})\big) \nonumber \IEEEeqnarraynumspace \\ 
 &-& (1-\alpha_u) p_i (x_i-x_i^b) (k_{u1}-k_{u0}) \big]- \widetilde{U}^{b}. \label{eq:V-V0}
\end{IEEEeqnarray}
\normalsize
\noindent The inequality in~\eqref{ccrcache:capacity_con} is the cache capacity constraint and, as expressed in~\eqref{ccrcache:continuous_x_y}, the control variables are continuous. Finally, the inequalities $U(Y) -U^{b} \geq0$  and  $V(Y, X) -V^{b} \geq 0$ are implicit domain constraints.

\theoremstyle{plain} \newtheorem{lemma:ccrcache}[lemma:biconvex]{Lemma} 
\begin{lemma:ccrcache} \label{lemma:ccrcache}
The CCRCache Problem is bi-convex.
\end{lemma:ccrcache}

\begin{proof}
The objective function is bi-convex, \emph{i.e.,} convex with respect to $Y$ for every fixed $X$ and convex with respect to $X$ for every fixed $Y$, since the logarithm is a concave function, the utility function $U$ is linear with respect to $Y$, and the function $V$ is bilinear in $X$ and $Y$ (since it contains the products $y_{ui} x_i$, see~\eqref{eq:V-V0}). Furthermore, all the problem's constraints are linear.
\end{proof}

An approach to tackle a bi-convex optimization problem would be to transform it into an equivalent problem that is instead convex in $(X,Y)$. However, such transformations leading to convex equivalent problems are the exception, rather the rule.   Standard transformation ``tricks'' include replacing the products $y_{ui} x_i$ by new variables or discretizing  one of the variables involved in the product\cite{wei2020tutorials}. The former option is not possible in our problem~(since the variables $y_{ui}$   and $x_i$ appear also outside of this product), and the latter could lead to a problem with a large number of new variables. Another approach includes the GOP~(global optimization) algorithm that guarantees convergence to the global optimum~\cite{floudas1990global},~\cite{floudas2013deterministic}. Unfortunately, this algorithm comes at the cost of high complexity that could be prohibitive in real-world systems with vast catalogs and multiple users. 

Moving away from ``exact'' methods that attempt to find global optima, Alternate Convex Search~\cite{wendell1976minimization}, and more recently ADMM methods~\cite{boyd2011distributed}, have been popular heuristics for bi-convex problems. Although there are problem instances whose structure permits such algorithms to (provably) converge to global optima~(\emph{e.g.}, the well-known matrix factorization problem), they~(at best) guarantee convergence to stationary points.
 We saw in Sec.~\ref{subsec:distributed} how ADMM can be applied in order to provide a distributed solution. The same method can be applied for bi-convex problems since its core idea consists of splitting the main problem into subproblems.  Here, the  CCRCache Problem can be broken into a subproblem that contains the recommendation variables and another that contains the caching variables. In order to apply ADMM, we reformulate the CCRCache Problem into an equivalent problem by introducing new variables and adding  bilinear constraints:

\theoremstyle{definition} \newtheorem*{CCRcache_prime}{CCRCache${}^\prime$ Problem}
\begin{CCRcache_prime} 
\begin{IEEEeqnarray}{rcl} 
&\underset{X,Y, Z}{\text{ min }}&  \left[-\log \left(U(Y) - U^{b}\right) - \log\left( G(X,Y,Z) - V^{b} \right)\right] \label{ccrcacheprime:obj} \IEEEeqnarraynumspace \\
&\text{s.t.}&  \eqref{ccrcache:N_recomm}, \eqref{ccrcache:capacity_con}, \nonumber\\
& & z_{ui}= x_i y_{ui}, \;\forall u\in\mathcal{U}, i\in\mathcal{K},  \label{ccrcacheprime:z}\\
& & x_{i}, y_{ui}, z_{ui}\in[0,1] \label{ccrcacheprime:continuous},
\end{IEEEeqnarray}
\noindent where the $Z=(z_{ui}\in [0,1])$ are auxiliary variables that replace the products $x_i y_{ui}$ and $G(X,Y,Z)$ is defined as follows:
\end{CCRcache_prime}
\begin{IEEEeqnarray}{rcl} 
G(X,Y,Z)&=&\sum_{u, i} \Big[\frac{\alpha_u}{N_u} \left(y_{ui}( \Lambda_{ui}   - k_{u0}) - z_{ui} (k_{u1}-k_{u0})\right) \nonumber \IEEEeqnarraynumspace \\ 
 &+& (1-\alpha_u) p_i (\lambda - k_{u0} - x_i (k_{u1}-k_{u0})) \Big].
\end{IEEEeqnarray}

It is important to note that the objective of the CCRCache${}^\prime$ Problem is convex in $(X,Y,Z)$ while the bi-linear constraints in~\eqref{ccrcacheprime:z} couple all variables together. We describe below how ADMM~\cite[Sec.~9.2]{boyd2011distributed} can be applied in the CCRCache${}^\prime$ Problem. Even though ADMM for bi-convex problems has no guarantee of convergence, it is expected to have better convergence properties~(faster convergence to a local or global optimum or better objective function value) than other local heuristics~\cite{boyd2011distributed}.

\theoremstyle{definition} \newtheorem*{CCRCache_alg_def}{The CCRCache algorithm}
\begin{CCRCache_alg_def}
The CP and the CDN exchange the following information: $\alpha_u$, $N_u$, $Y^{b}$, $U$, $U^{b}$, $\rho$, $G$, and $V^{b}$. 
Then, the two entities together~(or through a mediator) solve iteratively the CCRCache${}^\prime$ problem. At every iteration $k+1$ and for penalty parameter $q$, the following steps take place:
\begin{itemize}
\item Solving the $(Y,Z)$-subproblem: 
\footnotesize
\begin{IEEEeqnarray}{l} 
\left(Y^{(k+1)}, Z^{(k+1)}\right) = \argmin_{\text{s.t. \eqref{ccrcache:N_recomm},\eqref{ccrcacheprime:continuous}}} \bigg[ -\log\left(U(Y)-U^{b}\right)\nonumber \IEEEeqnarraynumspace\ \\
- \log\left(G(X^{(k)}, Y, Z) -V^{b}\right) \nonumber \\+ \frac{q}{2} \Big|\Big| Z- \left(diag(X^{(k)}) Y\right)^T + H^{(k)}\Big|\Big|_{F}^2 \bigg]. \IEEEeqnarraynumspace \label{eq:yzproblem} 
\end{IEEEeqnarray}
\normalsize
\item Solving the $X$-subproblem:
\footnotesize
\begin{IEEEeqnarray}{rcl}
 X^{(k+1)} &=& \argmin_{\text{s.t. \eqref{ccrcache:capacity_con},\eqref{ccrcacheprime:continuous}}} \bigg[ -\log\left(G(X, Y^{(k+1)}, Z^{(k+1)}) -V^{b} \right)\nonumber \IEEEeqnarraynumspace\ \\
&+& \frac{q}{2} \Big|\Big| Z^{(k+1)}- \left(diag(X) Y^{(k+1)}\right)^T + H^{(k)} \Big|\Big|_{F}^2 \bigg]. \IEEEeqnarraynumspace \label{eq:xproblem}
\end{IEEEeqnarray}
\normalsize

\item Updating the dual variables denoted by $H$:
\footnotesize
\begin{IEEEeqnarray}{rcl}
H^{(k+1)} &=& H^{(k)} + \left(Z^{(k+1)}- \left(diag(X^{(k+1)}) Y^{(k+1)}\right)^T \right). \IEEEeqnarraynumspace
\end{IEEEeqnarray}
\normalsize
\end{itemize}
\end{CCRCache_alg_def}

\noindent Essentially, the CCRCache algorithm splits the CCRCache${}^\prime$ problem into $(Y,Z)$- and $X$-subproblems that do not contain any coupling constraints. Both subproblems are   convex~(in fact strongly convex) and can be solved efficiently through standard interior-point or dual methods. The last terms of \eqref{eq:yzproblem} and \eqref{eq:xproblem} ensure that the coupling constraints in \eqref{ccrcacheprime:z} are not violated. Such violations are further controlled through the updates of the dual variables~(during the third step). 
The iterations can terminate according to standard residual criteria, \emph{i.e., }when the differences $z_{ui}-x_i y_{yi}$ are sufficiently small. 
As a final remark, we stress here that we do not claim that this is necessarily the best method for this problem, and other techniques could further enhance the method's performance~\cite{diamond2018general}. Our sole goal  is to apply a reasonably tested method for such problems, and evaluate if the control over the caching variables can reap additional benefits~(see Sec.~\ref{sec:perf}).

\section{Performance Evaluation}\label{sec:perf}

In this section, we evaluate numerically the payoffs that can be achieved through the proposed cooperation scheme. We will study two scenarios:  Scenario~I will focus on the evaluatation of the CCR and DCR algorithms in terms of cooperative gains and their impact on the quality of recommendations, while investigating the role of key problem parameters;  Scenario~II will focus on exploring the benefits of the CCRCache algorithm.
First, we present the default input parameters that, unless otherwise stated, will be used across the simulations.

\emph{Catalog and Recommendations: }
Our scenario consists of $100$ users who have access to  a catalog of $6000$ contents\footnote{According to \cite{netflix_museum}, in 2019, the total number of titles (movies and TV shows) available on Netflix in the USA was equal to $5848$.}, \emph{e.g.,} movies.  Without loss of generality, we consider equal-sized contents of
1Gb\footnote{Our performance evaluation results are similar in case of contents of heterogeneous sizes since the CP's revenues and the CDN's costs related to each content vary in a wide range.}. Every user receives $N_u=5$ recommendations and the probability of following the recommendations varies in $[0.6, 1)$, as in Netflix, where the average is equal to $0.8$~\cite{gomez2016netflix}.  For the matrix of content relevances $r_{ui}$, a subset of the Movielens dataset~\cite{movielens-related-dataset} containing 5-star ratings of movies was used. The ratings were mapped in the interval $[0,1]$ and we performed matrix completion to obtain the missing ratings~(as in~\cite{tsigkari2022approximation}).
Finally,  the baseline recommendations~(before any cooperation) for a user $u$, 
\emph{i.e.,} $y^{b}_{ui}$, are the $N_u$ contents that bring the highest revenue~($R_{ui}$) to the CP.

\emph{Caching Topology: } We consider a network of $9$ caches whose capacity will be specified in what follows and a root cache containing all contents. Every user has access to $2$ of the caches~(chosen randomly) and  to the root cache.
We assume that the (baseline) caching allocation, \emph{i.e.,} $X^b$, as decided by the CDN, is based on a popularity distribution over the catalog as observed by every cache in a time period that precedes the cooperation. For this, we set the content popularities observed by cache $j$ to be the normalized content utilities $r_{ui}$ aggregated over the connected users, \emph{i.e.}, $r_{ui}/\sum_{u\in \mathscr{C}_j} r_{ui}$, where $\mathscr{C}_j$ is the set of connected users to the cache $j$.

\emph{Revenues and Costs: } Based on the subscription prices of a major streaming platform in U.S.A.~\cite{netflix_price_usa}, the average time a user spends on the platform~\cite{variety_two_hours}, and the average length of movies~\cite{avg_length_movie}, we estimate that a user pays an approximate price of $\$0.36$ per movie. Taking into account licensing or production costs, we estimate that  $R_{ui}$~(CP's revenue per content) varies from $\$0.15$ to $\$0.24$ per movie\footnote{The  licensing and production costs we considered are realistic and are based on estimations for a movie of Netflix production. In particular, according to publicly available data on views count~\cite{variety_bird_box} and on production budget~\cite{wikipedia_bird_box}, the Netflix's film ``Bird Box'' had an approximate production cost of $\$0.2$ per view, as per 2020.}. The values of $R_{ui}$ were derived  through an equation that depends on the content relevances~(see Sec.~\ref{subsec:revenues_costs}), and, unless otherwise stated, this equation will be: $R_{ui} = 0.15 + 0.09 r_{ui}$~(in $\$$). This could, for example, capture an ad-based revenue model where $r_{ui}$ can be interpreted as the  user retention rate and, thus, the quantity $0.09 r_{ui}$ is the portion of ad-based revenue. Alternatively, this could also reflect a subscription-based revenue model where the licensing costs depend on the watched portion of the movie. 
Therefore, the baseline recommendations $Y^{b}$ are the ones with the highest relevances $r_{ui}$ per user. Finally, it is worth noting that, in~\cite{tsigkari2021globecom}, our numerical evaluations were performed for $R_{ui}$ being a concave function of $r_{ui}$ that could similarly capture an ad-based, subscription-based, or hybrid revenue model.  Under this assumption, we obtained similar performance results as the ones presented here.

The CDN charges the CP  $\$0.11$ per Gb~(according to~\cite{cloudfront_pricing} for the delivery).  Concerning the CDN's retrieval costs, they have been chosen randomly from the range $[0.0005, 0.02]$($\$$) for the connected caches, while the cost for the root cache is fixed at $\$0.055$. These values are in line with the simulation parameters used in related work on CDN's economics~\cite{gourdin2017economics}, where retrieval costs~(from caches nearby or origin server) vary in a wide range between $0.1\%$ and $50\%$ of the delivery fees  the CDN charges.

\subsection{Scenario I}
\begin{figure}[ht] 
  \centering

  \begin{tabular}{@{}c@{}}

    \includegraphics[width=7.7cm, trim={0.35cm 0.39cm 1cm 0.95cm},clip]{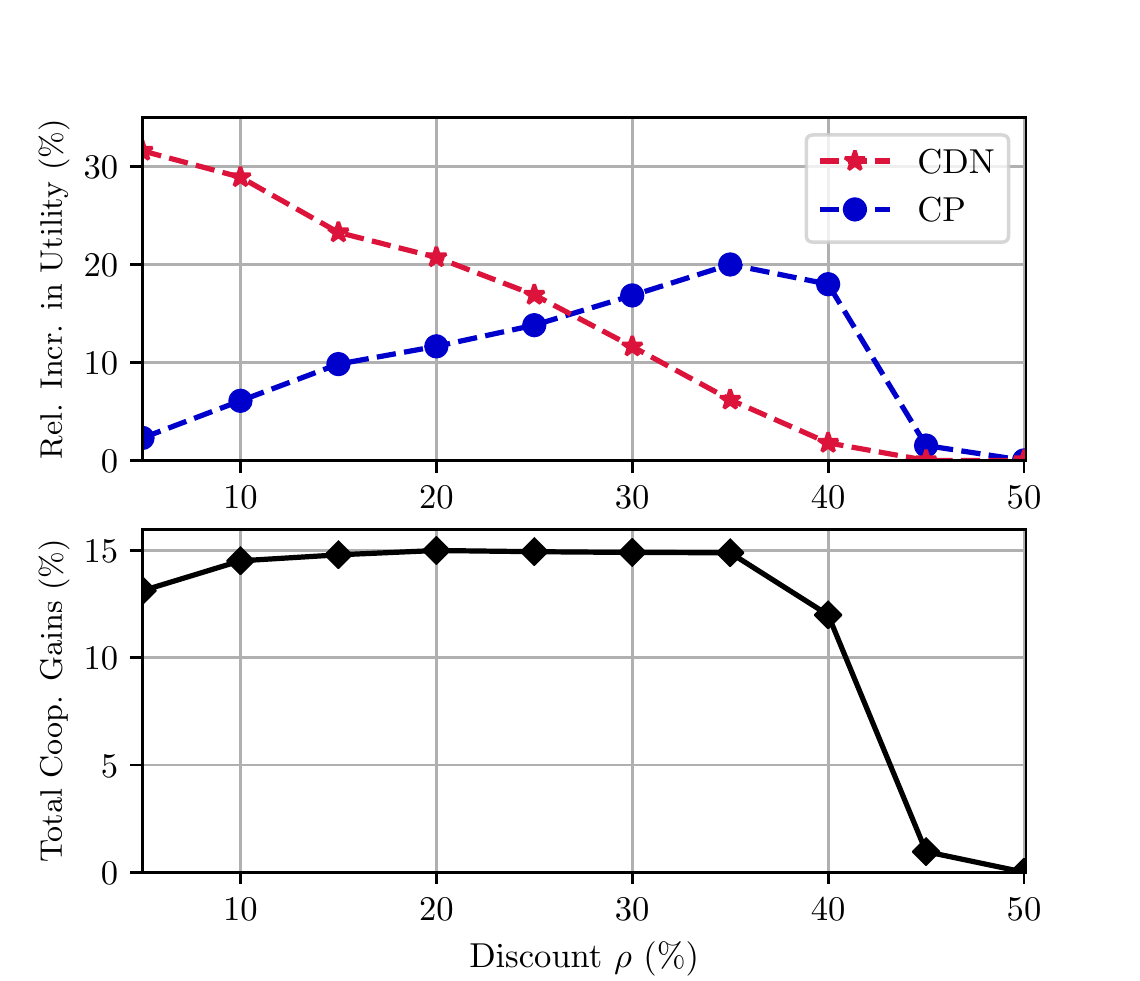} 
  \end{tabular}
\vspace{-0.23cm}

  \begin{tabular}{@{}c@{}}

    \includegraphics[width=7.7cm, trim={0.1cm 0cm 0.95cm 0.9cm},clip]{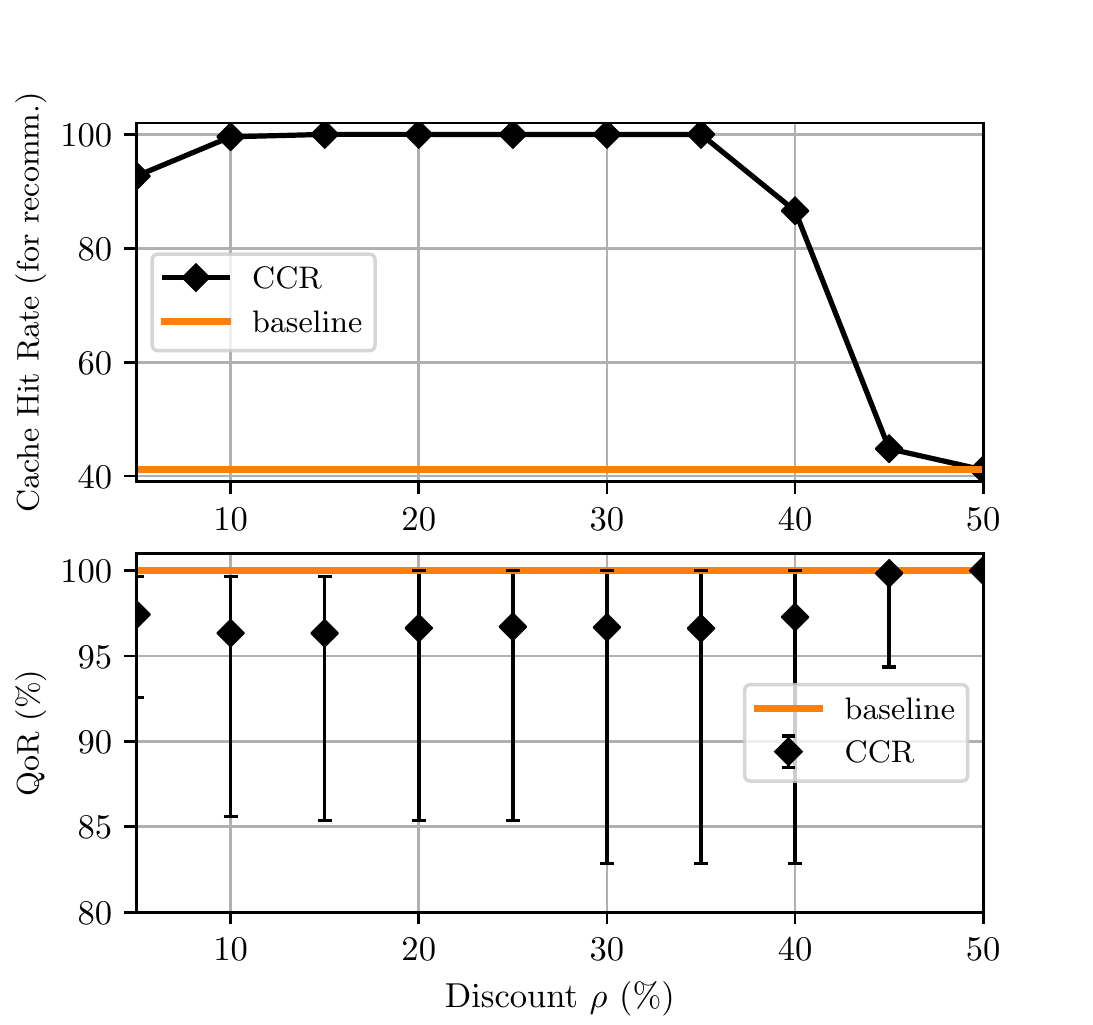}
  \end{tabular}

  \caption{Scenario I: Relative increase in utility for the CP and the CDN, total cooperative gains, cache hit rate for recommendations, and quality of recommendations achieved through the proposed cooperation scheme~(CCR algorithm) for different values of the discount $\rho\in[0.05, 0.5]$.} \label{fig:4in1}
\end{figure}

For the default parameters that were described above, for cache sizes varying~(randomly) from $1-4\%$ of the content catalog, and for different values of the discount $\rho$, we evaluate the proposed cooperation in Fig.~\ref{fig:4in1}.

The first subplot~(top) depicts the relative gains in utility for the two entities, \emph{i.e.,} the quantities $100 \cdot (U-U^{b})/U^{b}$ and $100 \cdot (\widetilde{U}-\widetilde{U}^{b})/\widetilde{U}^{b}$, as given by the CCR algorithm. We observe that, for low discount, the CDN benefits from the cooperation more than the CP. This is because the CDN saves on routing costs without its revenue from the delivery fees decreasing significantly. On the other hand, we see that the CP benefits the most for high discount as its savings on the delivery fees become important. However, for very high discount, close to $50\%$, the cooperation becomes unprofitable for the CDN and, therefore, the cooperation would cease, and the recommendations would revert back to the baseline ones. 
 It is important to highlight here that these points are Pareto optimal points. As explained in Sec.~\ref{sec:problem}, this means there is no other 
solution that is better than the solution for one entity and not
worse than the solution for the other entity. In the second subplot~(of Fig.~\ref{fig:4in1}), we plot the total relative  gains achieved from the cooperation, \emph{i.e.,} the quantity $(U-U^{b}+\widetilde{U}-\widetilde{U}^{b})/ (U^{b} + \widetilde{U}^{b})$.

 \theoremstyle{definition} \newtheorem{obs:cooperation_gains}{Observation} 
\begin{obs:cooperation_gains}
 The proposed cooperation  can lead to significant gains, up to $32\%$ for the CDN and up to $20\%$ for the CP in our scenario. The total cooperative gains can reach up to $15\%$ when compared to the total baseline utilities.
 \end{obs:cooperation_gains}

 It is worth noting that even gains of $3\%$ or $6\%$~(\emph{i.e.,} CP's gains for $\rho=5\%$ and $10\%$ respectively) already correspond to very large absolute monetary sums saved~(if one extrapolates to a much larger pool of users and requests, as in practice). Especially when referring to large CPs, like Netflix, that report annual profits of more than $2$ billion US dollars~\cite{fortune_netflix}.

Even though each pair of points in the top subplot corresponding to a value of $\rho$ is Pareto optimal, we see that  $\rho$ affects the gains of each entity. Obviously, the CDN would rather offer only a small discount, while the CP would prefer the largest discount possible. One could argue that the ``best'' $\rho$ is between $25\%$ and $30\%$, \emph{i.e., } where the two lines meet, since it does not give advantage to any entity. Defining what is the ``best'' $\rho$ and devising a method to find it is an interesting direction for future work. For example, one could model it as a game with alternating offers, or simply determine $\rho$ through exhaustive search from this plot. Besides, this plot reveals the effect of possible regulatory interventions that, \emph{e.g.,} could set bounds on such discounts in order to foster new business models, protect users' interests and so on.

In the third subplot~(of Fig.~\ref{fig:4in1}), we depict the cache hit rate for the small caches generated by the cooperative recommendations, \emph{i.e.,} the quantity $\sum_{u,i} \sum_{j\in\mathcal{C}(u) \setminus C_0} \alpha_u /N_u y_{ui}x_{ij}$, where $\mathcal{C}(u)\setminus C_0$ is the set of small caches that user $u$ is connected to. Note that $\alpha_u /N_u$ is the probability  the user will click on a specific recommendation. We also plot the cache hit rate of the baseline recommendations $Y^{b}$. We see that, before cooperation, only $42\%$ of the recommendations were generating a cache hit at the CDN's caches while this percentage can go up to $100\%$ for the cooperative recommendations. We remind the reader that, in our scenario, every user is connected to two small caches, and, therefore, we count a cache hit when the content in question is cached in at least one of the two caches. In fact, the cache hit rate could be smaller in scenarios where every user is connected to a single cache, or where the baseline caching allocation contains less popular/relevant contents. More importantly, 
 even if the CDN can serve from its small caches a big portion of the requests that come from recommendations, there is still room for improvement: these cache hits are not necessarily at the caches closest to each user. We will elaborate on that in Sec.~\ref{subsec:scenarioII}~(Scenario~II). We stress here that high cache hit rate is also beneficial for the user since it implies small start-up delays.

 In the forth subplot~(of Fig.~\ref{fig:4in1}), we investigate the impact of cooperative recommendations on the users' perception of the recommender. For that, we measure the quality of recommendations~(QoR) as defined in~\cite{tsigkari2022approximation}. In particular, QoR for user $u$ measures the sum of relevance of the received recommendations: $\sum_i r_{ui}y_{ui}$. The forth subplot shows the aggregate QoR (summed over the users) achieved by the cooperative and the baseline recommendations. The y-axis is regularized with respect to the highest existing relevances. The errorbars show the minimum and maximum QoR observed for individual users for every instance.

\theoremstyle{definition} \newtheorem{obs:QoR}[obs:cooperation_gains]{Observation} 
\begin{obs:QoR}
As the cooperative recommendations favor cached items and significantly increase the cache hit rate, the users' aggregate QoR is barely compromised~($ \geq 96\%$) in our scenario. The user's QoR  is at least $83\%$, where $100\%$ stands for the most relevant recommendations and the baseline here.
\end{obs:QoR}

\begin{figure}[!t]
  \centering

    \includegraphics[width=7.84cm, trim={0.27cm 0.27cm 0.3cm 0.34cm},clip]{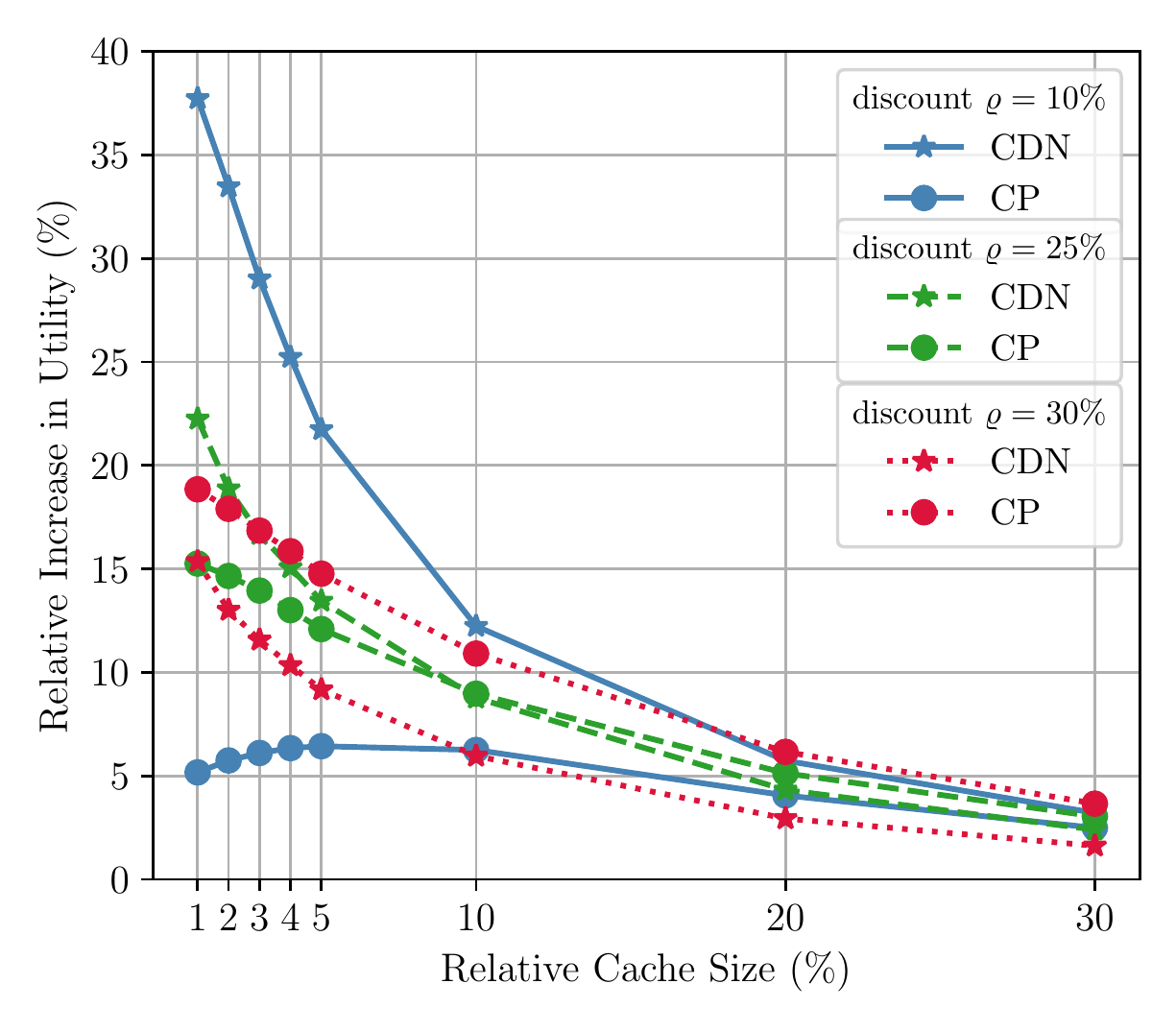} 
    \caption{Scenario I: Relative increase in utility for different discount values $\rho$ and for different relative cache sizes~($1-30\%$).} \label{fig:cache_size}
    \end{figure}

Next, we perform a sensitivity analysis with respect to two key problem parameters: the capacity of CDN's caches and the CP's revenues $R_{ui}$.  For the default simulation parameters, Fig.~\ref{fig:cache_size} depicts the relative increases in utility, as obtained by the CCR algorithm, for different relative cache sizes and different values of discount $\rho$.

\begin{figure}[!t]
  \centering
    \includegraphics[width=7.9cm, trim={0cm 0cm 0.27cm 0.27cm},clip]{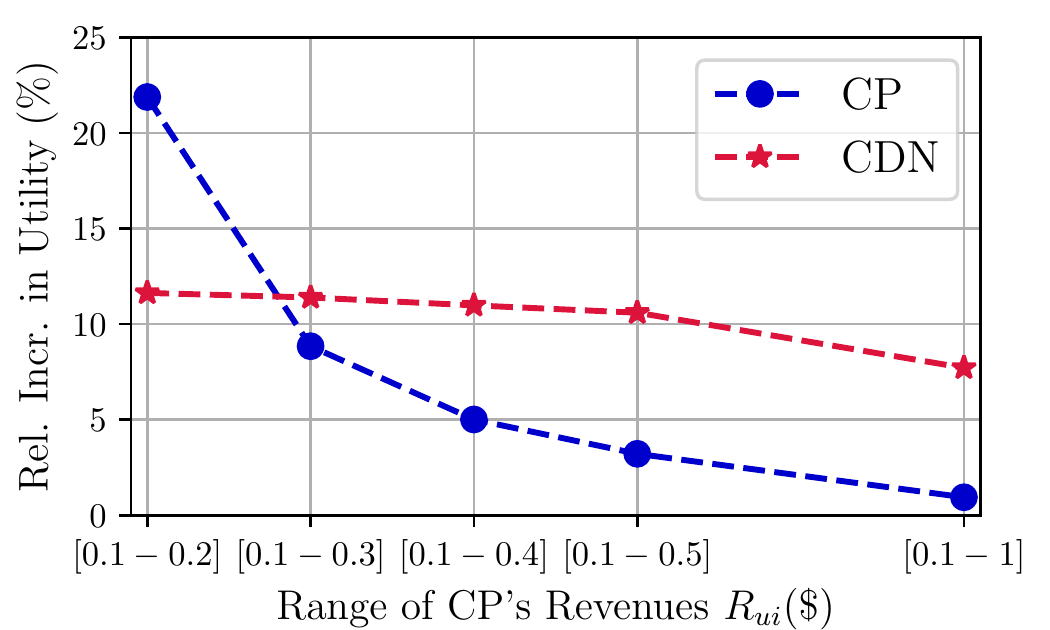} 
    \caption{Scenario I: Relative increase in utility for different ranges of $R_{ui}$~(CP's revenues per content) as obtained by the CCR algorithm.}\label{fig:rev_range}
\end{figure}

 \theoremstyle{definition} \newtheorem{obs:cache_size}[obs:cooperation_gains]{Observation} 
\begin{obs:cache_size}
As the relative cache size decreases, we notice the highest utility gains for both the CP and the CDN.
\end{obs:cache_size}

The observation above is particularly promising for today's and future wireless architectures where base stations are equipped with caches of small capacity. 
As the cache size increases~($10-30\%$ of the catalog), the utility gains decrease. Note that when the cache capacity is large, the baseline recommendations~($Y^{b}$) are likely to be already cached. Therefore, fewer~(when compared to the case of small cache capacity) recommendations need to be adjusted to favor cached items.

Next, we fix the discount at $30\%$ and the cache size at $1-4\%$. In Fig.~\ref{fig:rev_range}, we see how the CP's revenue values $R_{ui}$ affect the payoffs of the cooperation. 
We have plotted the relative increase in utility for both entities for $5$ different revenue ranges from $[0.1, 0.2)$ to $[0.1, 1)(\$$). We observe that, for the range $[0.1, 0.2)$, the CP could have an increase of $22\%$ of its utility. Then, as the range widens, the payoff for the CP decreases. In fact, when the  CP's  average revenue $R_{ui}$ is much larger than the delivery fee, a reduction on the fee will not have a significant impact on the CP's utility. On the other hand, the CDN's payoff is not affected as much as the range changes since its utility function does not contain the parameters $R_{ui}$.

\theoremstyle{definition} \newtheorem{obs:revenue_range}[obs:cooperation_gains]{Observation} 
\begin{obs:revenue_range}
When the range of $R_{ui}$ is narrow, the CP can enjoy an increase in its utility of $22\%$, for discount $\rho=30\%$. As the range widens, the CP would need a higher discount in order to keep the gains at the same level.
\end{obs:revenue_range}

\begin{figure}[!t]
\centering
\includegraphics[width=6.8cm, trim={0.86cm 0.85cm 0.42cm 0.5cm},clip]{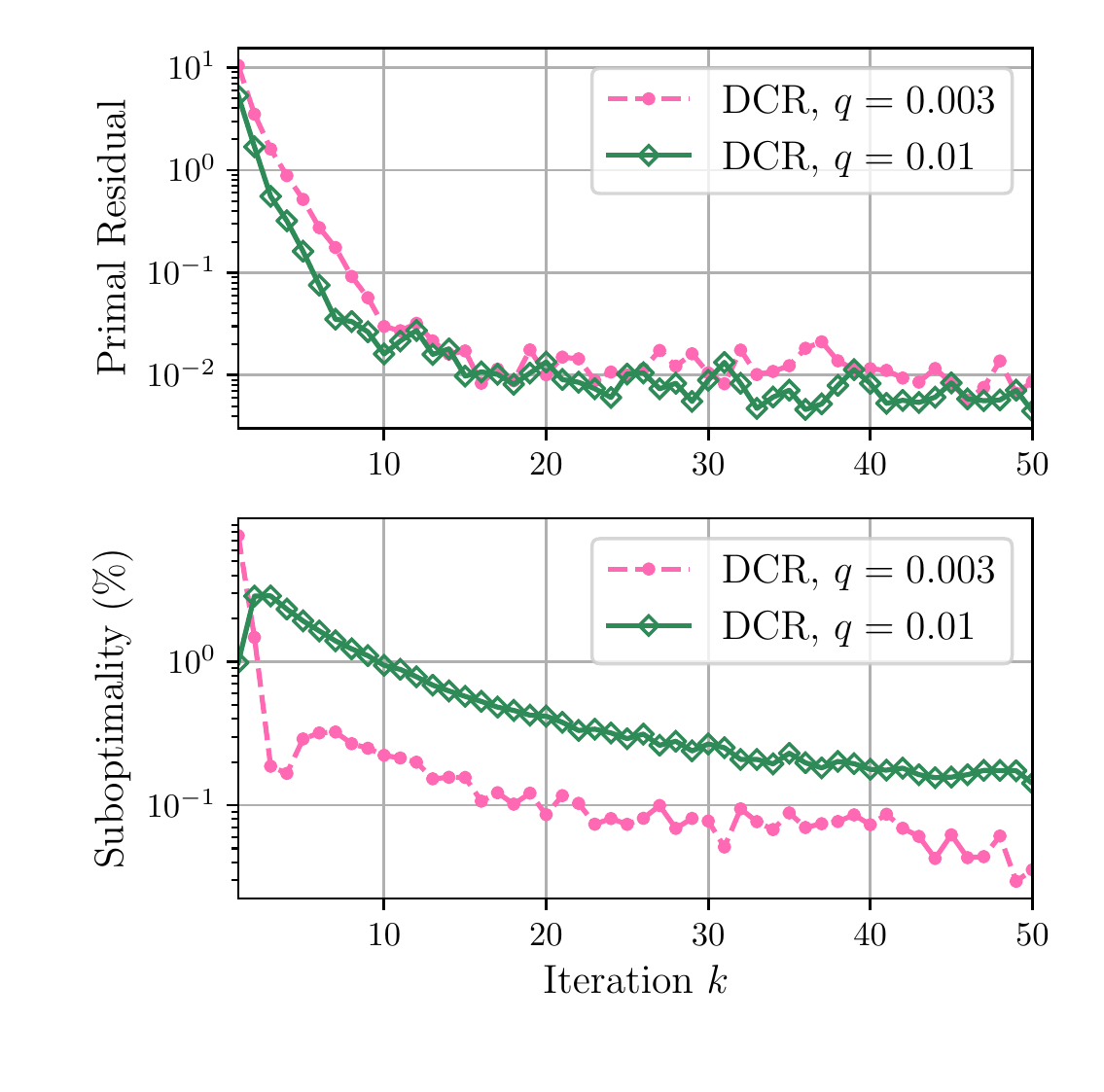}

\caption{Convergence of DCR algorithm in Scenario~I: Primal residual, \emph{i.e.,} $||\Psi^{(k)} - \widetilde{\Psi}^{(k)}||_F$, and suboptimality gap, \emph{i.e.,} $|DO^{(k)}-p^*|/|p^*|$, versus number of iterations for two different values of the penalty parameter $q$.}
\label{fig:suboptimality}
\end{figure}

In the remainder of this subsection, we will focus on the proposed distributed algorithm~(DCR). For the same problem parameters as in Fig.~\ref{fig:4in1} and the discount fixed at $30\%$, we will evaluate the convergence of the DCR algorithm and its impact on the cooperation payoffs. The top subplot of Fig.~\ref{fig:suboptimality} depicts the primal residual obtained within $50$ iterations for two different values of the penalty parameter $q$~(see eq.~\eqref{eq:update_dual_variables} in Sec.~\ref{subsec:distributed}). Note that the primal residual at iteration $k$ is equal to $||\Psi^{(k)} - \widetilde{\Psi}^{(k)}||_F$ and it measures how different the CP's and CDN's local solutions are. In the bottom subplot, we plot the suboptimality gap in percentage, \emph{i.e.}, $|DO^{(k)}-p^*|/|p^*|$, at iteration $k$, where $p^*$ is the optimal objective function value that is obtained by the CCR algorithm. This gap measures how far the distributed objective value is from  the centralized one and, according to Lemma~\ref{lemma:admm}, tends to zero for a sufficiently large number of iterations. Note that, as $p^*$ is in principle unknown, only the primal and dual residuals are used as stopping criteria. 

As we know from the theory on ADMM~\cite{boyd2011distributed}, the  higher the penalty parameter is, the lower the primal residuals are. In fact, for $q=0.01$,  we observe a residual's value of less than $4\cdot 10^{-3}$ and suboptimality gap of $0.14\%$. On the other hand, when $q=0.003$, the residual and the suboptimality gap are equal to $6\cdot 10^{-3}$ and $0.03\%$ respectively. These numbers show a rather fast convergence for the size of our scenario. However, this performance can be further enhanced by applying techniques that, although do not guarantee faster convergence, can work well in practice~(see~\cite{boyd2011distributed} for a review on such techniques).

\theoremstyle{definition} \newtheorem{obs:iteration}[obs:cooperation_gains]{Observation} 
\begin{obs:iteration}
Within only 3 iterations, the~(distributed) DCR algorithm can reach a suboptimality gap of less than $1\%$ when compared to the optimal objective function value achieved by the~(centralized) CCR algorithm. 
Within  $20$ iterations, the suboptimality gap is less than $0.1\%$. 
\end{obs:iteration}

\begin{table}[hbt] 
\centering 

\caption{Relative gains obtained by DCR${}^*$ and CCR} 
\begin{tabular}{|c|c|c|c|c|}
\hline
 & DCR & DCR & DCR& \multirow{2}{*}{CCR} \\
 & $k=2$ & $k=15$ & $k=30$&  \\
\hline
\textbf{CP's gains $(\%)$}  & $17.67$ & $16.79$ & $16.81$ & $16.84$ \\
\hline
\textbf{CDN's gains  $(\%)$} & $11.65$ & $11.63$ & $11.63$  & $11.63$  \\
\hline
\multicolumn{5}{l}{ \small{${}^*$}\scriptsize{$k$ stands for number of iterations of the DCR algorithm}}  
\end{tabular} \label{table:dcr}

\end{table}

Finally, Table~\ref{table:dcr} shows the CP's and CDN's relative gains that result from  the DCR algorithm~(with $q=0.003$) for different number of iterations and from the CCR algorithm.

\theoremstyle{definition} \newtheorem{obs:iteration2}[obs:cooperation_gains]{Observation} 
\begin{obs:iteration2}
Within only a few iterations, the relative increases in utility obtained by the the DCR algorithm approach the Pareto optimal points obtained  by the CCR algorithm.
\end{obs:iteration2}

\subsection{Scenario II}\label{subsec:scenarioII}

 As we saw in Fig.~\ref{fig:4in1}, the cooperative recommendations can lead to a high cache hit rate at the CDN's caches. Although this rate implies  significant savings for the CDN, the cache hits do not necessarily happen at the cache that generates the lowest retrieval cost~(when delivered to each user). What is more, if the CDN could cache another content, that is potentially more related to the ones in the baseline recommendations, then the CP could further increase its benefits, without actually increasing the cache hit  rate, per se. For this reason, we will evaluate now the potential benefits of extending the cooperation towards caching decisions, as we discussed in  Sec.~\ref{sec:extension}, where we proposed the CCRCache algorithm. 
 
  \begin{figure}[th]
\centering
\includegraphics[width=8.9cm, trim={0.26cm 0.35cm 0.36cm 0.37cm},clip]{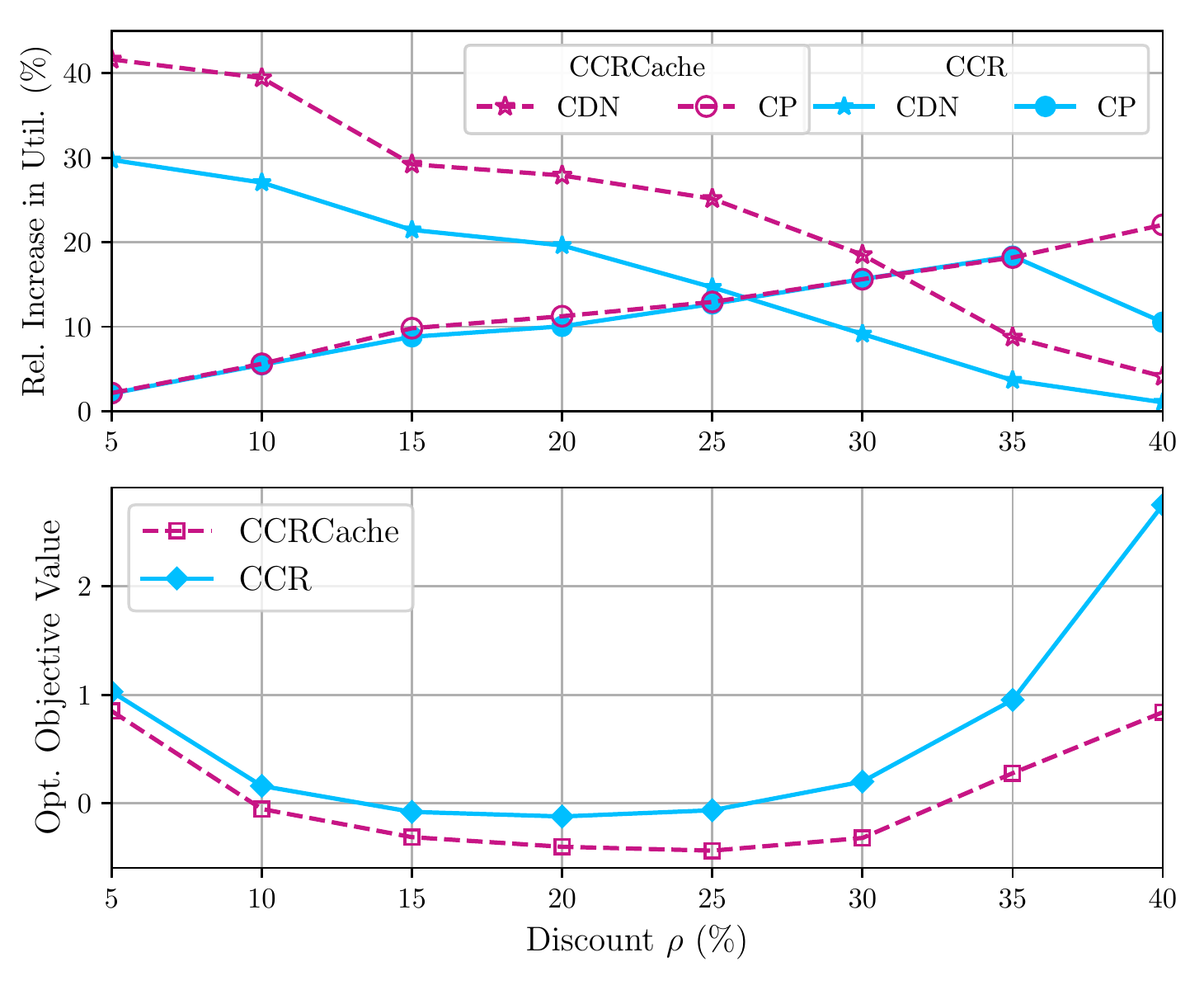}

\caption{Scenario II: Relative gains in utility~(for the two entities) and optimal objective function values achieved by the CCRCache and the CCR algorithm for different values of discount $\rho\in[0.05, 0.4]$.}
\label{fig:ccrcache}
\end{figure}

 For the default parameters that were described in the beginning of Sec.~\ref{sec:perf}, for capacity of caches equal to $5\%$ of the content catalog, and for different values of the discount $\rho$, we compare the CCR and CCRCache algorithms in Fig.~\ref{fig:ccrcache}. More specifically, we apply the CCR algorithm for every problem instance where only the recommendations are the cooperation variables and we apply the CCRCache algorithm for the same instance where caching is also a cooperation variable.
The top subplot depicts the relative gains in utility for the two entities, while the bottom subplot shows the objective function values obtained. We notice that CCRCache leads to larger gains~(for at least the CDN) and smaller~(better) objective functions values than the ones obtained by the CCR algorithm.

\theoremstyle{definition} \newtheorem{obs:ccrcache}[obs:cooperation_gains]{Observation} 
\begin{obs:ccrcache}
When caching becomes a cooperation variable, the CP-CDN cooperation, through the CCRCache algorithm, can boost CDN's utility up to $42\%$. At the same time, the CP's utility gains are at least as high as when recommendations are the only cooperation variables~(through the CCR algorithm).
\end{obs:ccrcache}

Finally, we  compare the CCRCache scheme with the related work on the joint optimization of caching and recommendations, \emph{e.g.,}~\cite{chatzieleftheriou2019joint-journal, tsigkari2022approximation, qi2018optimizing}. In contrast to our network-economical approach, the aforementioned works focus on maximizing network-related measures, such as cache hit rate. For this reason, we adapt these schemes towards  profit maximization.  Therefore, the state-of-the-art schemes could be  formulated as the problem of maximizing the aggregate profit, \emph{i.e.,} 

\begin{IEEEeqnarray}{lcl} 
&\underset{Y, X}{\text{ max }}& U(Y)+ V(X, Y) \label{eq:profit_max} \IEEEeqnarraynumspace \\ 
&\text{s.t.  }& \eqref{ccrcache:N_recomm}, \eqref{ccrcache:capacity_con}, \text{ and } \eqref{ccrcache:continuous_x_y} \nonumber,
\end{IEEEeqnarray}
\noindent where $U(Y)+ V(X, Y)=\sum_{u, i} [\frac{\alpha_u}{N_u}  y_{ui}  (R_{ui} - K_{ui}(X))+(1-\alpha_u) p_i (\lambda-K_{ui}(X))].$ We stress here that this formulation~(different to the NBS formulation we employed in the CCR and CCRCache problems) does not contain the baseline utilities. Furthermore, for the requests coming for recommendations,   the term of the CP's costs and the term of the CDN's revenues are canceled out in the sum of the two entities' utility functions and, for the requests outside of recommendations,  the term $\Lambda_{ui}$ in CDN's revenues is replaced by $\lambda$ since no discount on the delivery fees applies for the CP, as is the case in related work. 

Table \ref{table:versus_related_work} shows the relative gains/losses in utility when solving the problem in \eqref{eq:profit_max} (which represents a profit-oriented joint caching and recommendation schemes like the ones in the literature) and when  applying the proposed  CCRCache algorithm for  $\rho=20\%$. We see that the former leads to a loss in profit for the CP~($-2\%$) and a gain in profit for the CDN~($+53\%$), as it was also the case in the toy example in Sec.~\ref{subsec:toy}. The CP's loss in profit is a result of recommending cached contents whose aggregate popularity is high, but they are not necessarily the most relevant to each user. We remind the reader that no discount on the delivery fees  applies in the problem  in~\eqref{eq:profit_max}.  On the other hand, our scheme provides incentives to the CP, it leads to gains in profit for both stakeholders~($+12\%$ and $+28\%$ respectively), and a proportional fair allocation of the gains~(as guaranteed by the NBS formulation).

\begin{table}[ht] 
\centering 

\caption{Proposed Cooperation versus Related Work: \\relative gains/losses in utility }
\begin{tabular}{|c|c|c|}
\hline
 & \multirow{2}{*}{\textbf{CP}} & \multirow{2}{*}{\textbf{CDN}}  \\
 & & \\
\hline
\textbf{Joint Caching and}  & \multirow{2}{*}{$-2\%$} &\multirow{2}{*}{$+53\%$} \\
\textbf{Recommendations Scheme${}^*$} &  &  \\
\hline
\textbf{Our Cooperation Scheme} & \multirow{2}{*}{$+12\%$} &\multirow{2}{*}{$+28\%$} \\
\textbf{CCRCache  for $\rho=20\%$}  &  &  \\
\hline
\multicolumn{3}{l}{ \small{${}^*$}\scriptsize{Adaptation of related work, \emph{e.g.,} \cite{chatzieleftheriou2019joint-journal, tsigkari2022approximation},}} \\
\multicolumn{3}{l}{\;\scriptsize{ towards aggregate profit maximization} }
\end{tabular} \label{table:versus_related_work}

\end{table}

\theoremstyle{definition} \newtheorem{obs:ccrcacheVSrelated}[obs:cooperation_gains]{Observation} 
\begin{obs:ccrcacheVSrelated}
In contrast to the state-of-the-art schemes for joint caching and recommendation, the proposed cooperation scheme~(CCRCache) provides concrete incentives to the CP to cooperate with the CDN  so that they  design together the caching and recommendation decisions.
\end{obs:ccrcacheVSrelated}

\section{Related Work}
Several works in literature focus on the cache-friendly recommendations or the joint caching-recommendation paradigm. In~\cite{cache-centric-video-recommendation}, the authors propose a reordering of the videos appearing in YouTube's related videos section by ``pushing'' on top of the list the cached items. Similar in spirit, \cite{kastanakis2020network} presents a method of replacing or reordering contents in the related videos section taking into account network-related costs or QoS metrics. A decomposition algorithm for the joint caching and recommendation problem is  proposed in~\cite{chatzieleftheriou2019joint-journal}. Targeting cache hit rate maximization, their policy first decides on caching, accounting for the impact of recommendations, and then adjusts the recommendations in order to favor cached items. In~\cite{tsigkari2022approximation}, the authors formulate the joint problem as a maximization of a user-centric metric consisting of expected QoS and quality of recommendations. The authors  propose an algorithm with approximation guarantees for this joint problem. Finally, in \cite{liu2019deep}, the authors  employ machine learning techniques to devise caching and recommendation policies taking into account the fetching cost of the content requests.
Since most of the works above assume that the same entity decides on caching and recommendations, they do not explore the financial aspects of the recommendations from the point of view of both the CP and the CDN. More importantly, none of the existing algorithms  guarantee a fair split of the financial gains that come from cache-friendly recommendations.

The theoretical framework of the NBS  that we employ in this work was introduced by John Nash in 1950 in~\cite{nash1950bargaining}. The NBS is a cooperation mechanism that has been employed, among others, in problems of spectrum access coordination~\cite{wu2013cooperative}, bandwidth allocation~\cite{yuan2017cooperative}, and content caching~\cite{wang2017milking}. More specifically, in~\cite{wang2017milking},   caches that belong to a network collaborate with each other in order to decide on the caching allocation. Moreover, in~\cite{jiang2009cooperative}, the authors model a CDN-ISP collaboration as a NBS problem.

Game theory has also been employed by works that study the dynamics between CPs and edge caching providers and propose cooperations or coalitions. For example, \cite{douros2017caching} and
~\cite{mitra2019consortiums} model a coalitional game  between a last-mile ISP and CPs. 
The authors in~\cite{ahmadi2020cache} suggest that the caching network providers should give incentives to the CP in a form of a subsidy~(that is paid in proportion to the savings that come from caching).
Nevertheless, these works focus on the caching allocation or deployment without exploiting the impact of recommendations on content requests.

This paper extends our earlier work~\cite{tsigkari2021globecom} by providing in-depth insights on the proposed cooperation scheme, extending the cooperation mechanism to a distributed algorithm and presenting a comprehensive evaluation
of the proposed algorithms in a variety of scenarios and for
different input parameters. Finally, the current work discusses
a possible extension of the presented problem towards the CDN's caching decisions.

\section{Conclusions and Future Work}

In this work, we proposed a novel cooperation framework in which the CP and the CDN jointly decide on the recommendations in order to favor cached contents. The optimization problem of the cooperation was formulated in such a way that the cooperative recommendations lead to a fair and efficient allocation of financial gains between the two entities. We also developed a distributed algorithm when the two entities are not willing to share private information on their revenue/cost functions. Furthermore, we explored how this cooperation framework could be extended towards the CDN's caching decisions. Although this problem is harder to solve, it has the potential to further increase the cooperation gains.
 Our numerical evaluations show that, in realistic scenarios, the two entities can benefit of  an increase in their expected net revenue of up to $37\%$ and up to $42\%$ when caching is a cooperation variable.

The cooperation model presented in this work could be extended in several directions. For example, as we discussed in Sec.~\ref{sec:perf}, one could add the discount parameter $\rho$ as a control variable in the
cooperation problem. Another direction would be to include the users as players  that could potentially enjoy lower subscription fees when they receive cooperative recommendations that diverge from their tastes. Finally, it would be interesting to design a mechanism~(on top of the proposed cooperation) that can address trust/security issues that could occur, \emph{e.g.,} misreported gains or misinformation between the stakeholders.

\bibliographystyle{IEEEtran}
\bibliography{IEEEabrv,tsigkari_quid-pro-quo}

\vfill

\end{document}